\documentclass[12pt,preprint]{aastex}
\usepackage{times,pslatex}
\usepackage{amsmath,amsfonts,graphicx}
\usepackage{epsfig}
\usepackage{rotate}

\newcommand{\vRstarA}{0.776}
\newcommand{\eRstarA}{0.009}
\newcommand{\vRstarB}{0.484}
\newcommand{\eRstarB}{0.024}
\newcommand{\vRplanet}{4.347}
\newcommand{\eRplanet}{0.099}
\newcommand{\vaplanet}{0.355}
\newcommand{\eaplanet}{0.002}
\newcommand{\veplanet}{0.118}
\newcommand{\eeplanet}{0.002}

\newcommand{\vMstarA}{0.820}
\newcommand{\eMstarA}{0.015}
\newcommand{\vMstarB}{0.542}
\newcommand{\eMstarB}{0.008}

\newcommand{\kepler}{{\it Kepler}}
\newcommand{\kic}{{Kepler-413~}}
\newcommand{\kicb}{{Kepler-413b~}}

\newcommand{\Logg}{\mbox{$\log g$}}

\slugcomment{Accepted for publication in the Astrophysical Journal}
\received{2013, Oct}
\begin{document}

\title{Kepler-413b: a slightly misaligned, Neptune-size transiting circumbinary planet}
\shorttitle{Circumbinary Planet \kicb}

\author{
V.~B.~Kostov\altaffilmark{1,2,12},
P.~R.~McCullough\altaffilmark{1,2},
J.~A.~Carter\altaffilmark{3,13},
M.~Deleuil\altaffilmark{4},
R.~F.~D\'{\i}az\altaffilmark{4},
D.~C.~Fabrycky\altaffilmark{5},
G.~H\'{e}brard\altaffilmark{6,7},
T.~C. Hinse\altaffilmark{8,9},
T.~Mazeh\altaffilmark{10}
J.~A.~Orosz\altaffilmark{11},
Z.~I.~Tsvetanov\altaffilmark{1},
W.~F.~Welsh\altaffilmark{11}
}

\email{vkostov@pha.jhu.edu}

\altaffiltext{1}{Department of Physics and Astronomy, Johns Hopkins University, 3400 North Charles Street, Baltimore, MD 21218}
\altaffiltext{2}{Space Telescope Science Institute, 3700 San Martin Dr., Baltimore MD 21218}
\altaffiltext{3}{Harvard-Smithsonian Center for Astrophysics, 60 Garden Street, Cambridge, MA 02138}
\altaffiltext{4}{Laboratoire d'Astrophysique de Marseille, 38 rue Fr\'ed\'eric Joliot-Curie, 13388 Marseille cedex 13, France}
\altaffiltext{5}{Department of Astronomy and Astrophysics, University of Chicago, 5640 South Ellis Avenue, Chicago, IL 60637}
\altaffiltext{6}{Institut dÕAstrophysique de Paris, UMR 7095 CNRS, Universit\'{e} Pierre \& Marie Curie, 98bis boulevard Arago, 75014 Paris, France}
\altaffiltext{7}{Observatoire de Haute-Provence, Universit\'{e} dÕAix-Marseille \& CNRS, 04870 Saint Michel lÕObservatoire, France}
\altaffiltext{8}{Korea Astronomy
and Space Science Institute (KASI), Advanced Astronomy and Space Science Division, Daejeon 305-348, Republic of Korea}
\altaffiltext{9}{Armagh Observatory, College Hill, BT61 9DG Armagh, NI, UK}
\altaffiltext{10}{Department of Astronomy and Astrophysics, Tel Aviv University, 69978 Tel Aviv, Israel}
\altaffiltext{11}{Department of Astronomy, San Diego State University, 5500 Campanile Drive, San Diego, CA 92182}
\altaffiltext{12}{NASA Earth and Space Science Graduate Fellow}
\altaffiltext{13}{Hubble Fellow}

\begin{abstract}
We report the discovery of a transiting, $R_p = \vRplanet\pm\eRplanet R_\oplus$, circumbinary planet (CBP) orbiting the {\it \kepler} $K+M$ Eclipsing Binary (EB) system KIC 12351927 (Kepler-413) every $\sim66$ days on an eccentric orbit with $a_p = \vaplanet\pm\eaplanet AU$, $e_p = \veplanet\pm\eeplanet$. The two stars, with $M_A=\vMstarA\pm\eMstarA  M_{\odot}, R_A=\vRstarA\pm\eRstarA R_{\odot}$ and $M_B=\vMstarB\pm\eMstarB M_{\odot}, R_B = \vRstarB\pm\eRstarB R_{\odot}$ respectively revolve around each other every $10.11615\pm0.00001$ days on a nearly circular ($e_{EB} = 0.037\pm0.002$) orbit. The orbital plane of the EB is slightly inclined to the line of sight ($i_{EB}=87.33\pm0.06\arcdeg$) while that of the planet is inclined by $\sim2.5\arcdeg$ to the binary plane at the reference epoch. Orbital precession with a period of $\sim11$ years causes the inclination of the latter to the sky plane to continuously change. As a result, the planet often fails to transit the primary star at inferior conjunction, causing stretches of hundreds of days with no transits (corresponding to multiple planetary orbital periods). We predict that the next transit will not occur until 2020. The orbital configuration of the system places the planet slightly closer to its host stars than the inner edge of the extended habitable zone. Additionally, the orbital configuration of the system is such that the CBP may experience Cassini-States dynamics under the influence of the EB, in which the planet's obliquity precesses with a rate comparable to its orbital precession. Depending on the angular precession frequency of the CBP, it could potentially undergo obliquity fluctuations of dozens of degrees (and complex seasonal cycles) on precession timescales.
\end{abstract}


\include{variables}

\section{Introduction}
\label{sec:intro}

A mere two years ago, \cite{doyle2011} announced the discovery of the first transiting circumbinary planet (CBP), Kepler-16b. Six more transiting CBPs, including a multi-planet system, a CBP in the habitable zone, and a quadruple host stellar system, have been reported since \citep{Welsh2012, Orosz2012, Orosz2012b, kostov2013, Schwamb2013}. In comparison, the number of planetary candidates orbiting single stars is significantly larger $-$ three thousand and counting \citep{burke2013}. 

Extensive theoretical efforts spanning more than two decades have argued that planets can form around binary stars \citep{alex12, paar12,pn07, pn08a, pn08b, pn08c, pn13, mart13, marz13, mes12a, mes12b, mes13, raf13}. Simulations have shown that sub-Jupiter gas and ice giants should be common, and due to their formation and migration history should be located near the edge of the CB protoplanetary disk cavity. Indeed that is where most of the CBPs discovered by {\it \kepler} reside! Once formed, it has been shown that CBPs can have dynamically stable orbits beyond certain critical distance \citep{holman99}. This distance depends on the binary mass fraction and eccentricity and is typically a few binary separations. All discovered CBP are indeed close to the critical limit -- their orbits are only a few tens of percent longer than the critical separation necessary for stability \citep{welsh14}. Additionally, models of terrestrial planet formation in close binary systems ($a_{bin} < 0.4AU$) indicate that accretion around eccentric binaries typically produces more diverse and less populated planetary systems compared to those around circular binaries \citep{quin06}. In contrast, the location of the ice line in CB protoplanetary disks is expected to be interior to the critical stability limit for 80\% of wide, low-mass binary systems ($M_{bin} < 4M_{\odot}$) with $a_{bin} \sim 1AU$ \citep{clan13}. Thus, Clanton argues, formation of rocky planets in such systems may be problematic. The theoretical framework of formation and evolution of planets in multiple stellar systems demands additional observational support, to which our latest CBP discovery \kic contributes an important new insight. 

The configurations of six of the confirmed CBPs are such that they currently transit their parent stars every planetary orbit. \cite{doyle2011} note, however, that the tertiary (planet transits across the primary star) of Kepler-16b will cease after 2018, and the quaternary (planet transits across the secondary star) after 2014. The last transit of Kepler-35b was at BJD 2455965 \citep{Welsh2012}; it will start transiting again in a decade. As pointed out by \cite{schneider1994}, some CBP orbits may be sufficiently misaligned with respect to their host EB and hence precessing such that the above behavior may not be an exception. Additionally, \cite{fouc13} argue that circumbinary disks around sub-AU stellar binaries should be strongly aligned (mutual inclination $\theta \le 2\arcdeg$), in the absence of external perturbations by additional bodies (either during or after formation), whereas the disks and planets around wider binaries can potentially be misaligned ($\theta \ge 5\arcdeg$). \cite{fouc13} note that due to the turbulent environment of star formation, the rotational direction of the gas accreting onto the central proto-binary is in general not in the same direction as that of the central core. Their calculations show that the CB disk is twisted and warped under the gravitational influence of the binary. These features introduce a back-reaction torque onto the binary which, together with an additional torque from mass accretion, will likely align the CB protoplanetary disks and the host binary for close binaries but allow for misalignment in wider binaries. 

The observational consequence of slightly misaligned CBPs is that they may often fail to transit their host stars, resulting in a light curve exhibiting one or more consecutive tertiary transits followed by prolonged periods of time where no transits occur. This effect can be further amplified if the size of the semi-minor axis of the transited star projected upon the plane of the sky is large compared the star's radius. 

Such is the case of \kic (KIC 12351927), a $10.116146$-day Eclipsing Binary (EB) system. Its {\em Kepler} light curve exhibits a set of three planetary transits (separated by $\sim66$ days) followed by $\sim800$ days with no transits, followed by another group of five transits (again $\sim66$ days apart). We do not detect additional events $\sim66$ days (or integer multiples of) after the last transit. Our analysis shows that such peculiar behavior is indeed caused by a small misalignment and precession of the planetary orbit with respect to that of the binary star. 

Here we present our discovery and characterization of the CBP orbiting the EB \kic. This paper is organized as an iterative set of steps that we followed for the complete description of the circumbinary system. In Section \ref{sec:kepler} we describe our analysis of the \kepler~data, followed by our observations in Section \ref{sec:followup}. We present our analysis and results in Section \ref{sec:photodynamics}, discuss them in Section \ref{sec:discussion} and draw conclusions in Section \ref{sec:conclusions}. 
\section{{\em Kepler} Data}
\label{sec:kepler}

\subsection{{\em Kepler} Light Curve}
\label{sec:lc}

We extract the center times of the primary ($T_{prim}$) and secondary ($T_{sec}$) stellar eclipses, the normalized EB semi major axes ($a/R_A$), ($a/R_B$), the ratio of the stellar radii ($R_B/R_A$), and inclination ($i_b$) of the binary and the flux contribution of star B from the \kepler~light curve. Throughout this work, we refer to the primary star with a subscript {\em``A''}, to the secondary with a subscript {\em``B''}, and to the planet with a subscript {\em``p''}. We model the EB light curve of \kic with ELC \citep{Orosz2012}. 

The \kepler~data analysis pipeline \citep{jen10} uses a cosmic-ray detection procedure which introduces artificial brightening near the middle of the stellar eclipses of \kic (see also \citealt{Welsh2012}). The procedure flags and corrects for positive and negative spikes in the light curves. The rapidly changing stellar brightness during the eclipse and the comparable width between the detrending window used by the pipeline and the duration of the stellar eclipse misleads the procedure into erroneously interpreting the mid-eclipse data points as negative spikes. This leads to the unnecessary application of the cosmic ray correction to the mid-eclipse data points prior to the extraction of the light curve. The target pixel files, however, contain a column that stores the fluxes, aperture positions and times of each flagged cosmic ray event. To account for the anomalous cosmic ray rejection introduced by the pipeline, we add this column back to the flux column using fv (downloaded from the {\em Kepler Guest Observer} website) and then re-extract the corrected light curve using the {\it kepextract} package from PyKE \footnote{http://keplergo.arc.nasa.gov/ContributedSoftwarePyKEP.shtml} \citep{still_xxx,kinem12}. We note that our custom light curve extraction from the target pixel files for Quarters 1 through 14 introduces a known timing error of $\sim67$ sec in the reported times which we account for. 

\begin{figure}
\centering
\plotone{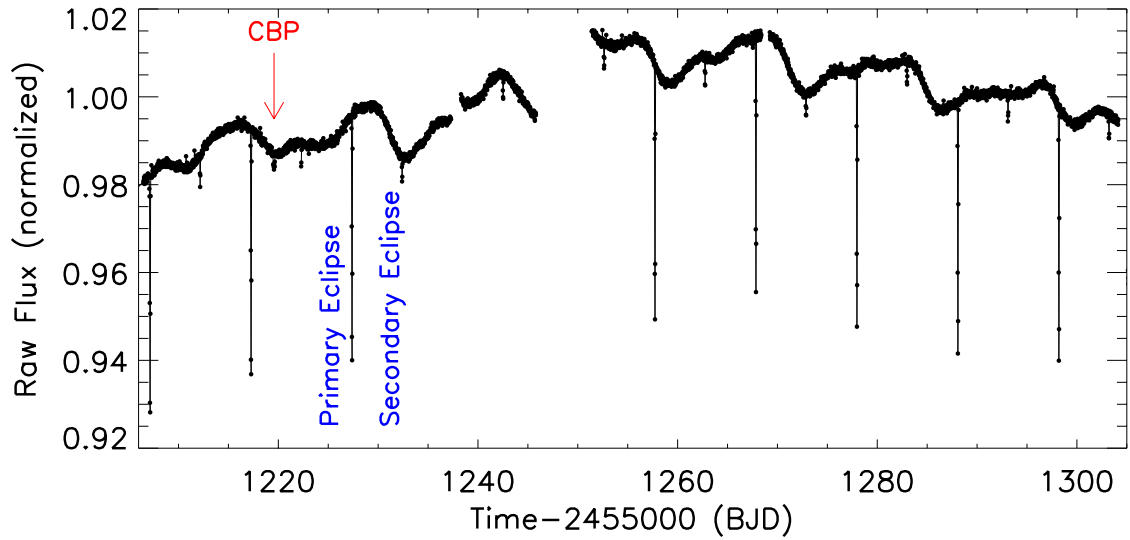}
\caption{
A section of the raw (SAPFLUX), normalized {\em Kepler} light curve of \kic spanning Quarter 15. The prominent stellar eclipses are clearly seen, with a depth of $\sim6\%$ and $\sim0.5\%$ for primary and secondary respectively. The last detected transit of the CBP is indicated with an arrow near day 1219. The gap near day 1250 is due to missing data. 
\label{fig:raw_lc}}
\end{figure}

Next, we detrend the normalized, raw {\em Kepler} data (SAPFLUX with a SAPQUALITY flag of 0) of \kic by an iterative fit with a high-order (50+) Legendre polynomial on a Quarter-by-Quarter basis. A representative section of the light curve, spanning Quarter 15 is shown in Figure \ref{fig:raw_lc}. We use a simple $\sigma$-clipping criteria, where points that are 3-$\sigma$ above and below the fit are removed and the fit is recalculated. Next, the stellar eclipses are clipped out. We note that for our search for transiting CBP we do this for the entire EB catalog listed in \cite{slawson11, kirk14}. The order of execution of the two steps (detrending and removal of stellar eclipses) generally depends on the baseline variability of the particular target. For quiet stars (like \kic) we first remove the eclipses and then detrend. 

\begin{figure}
\centering
\plotone{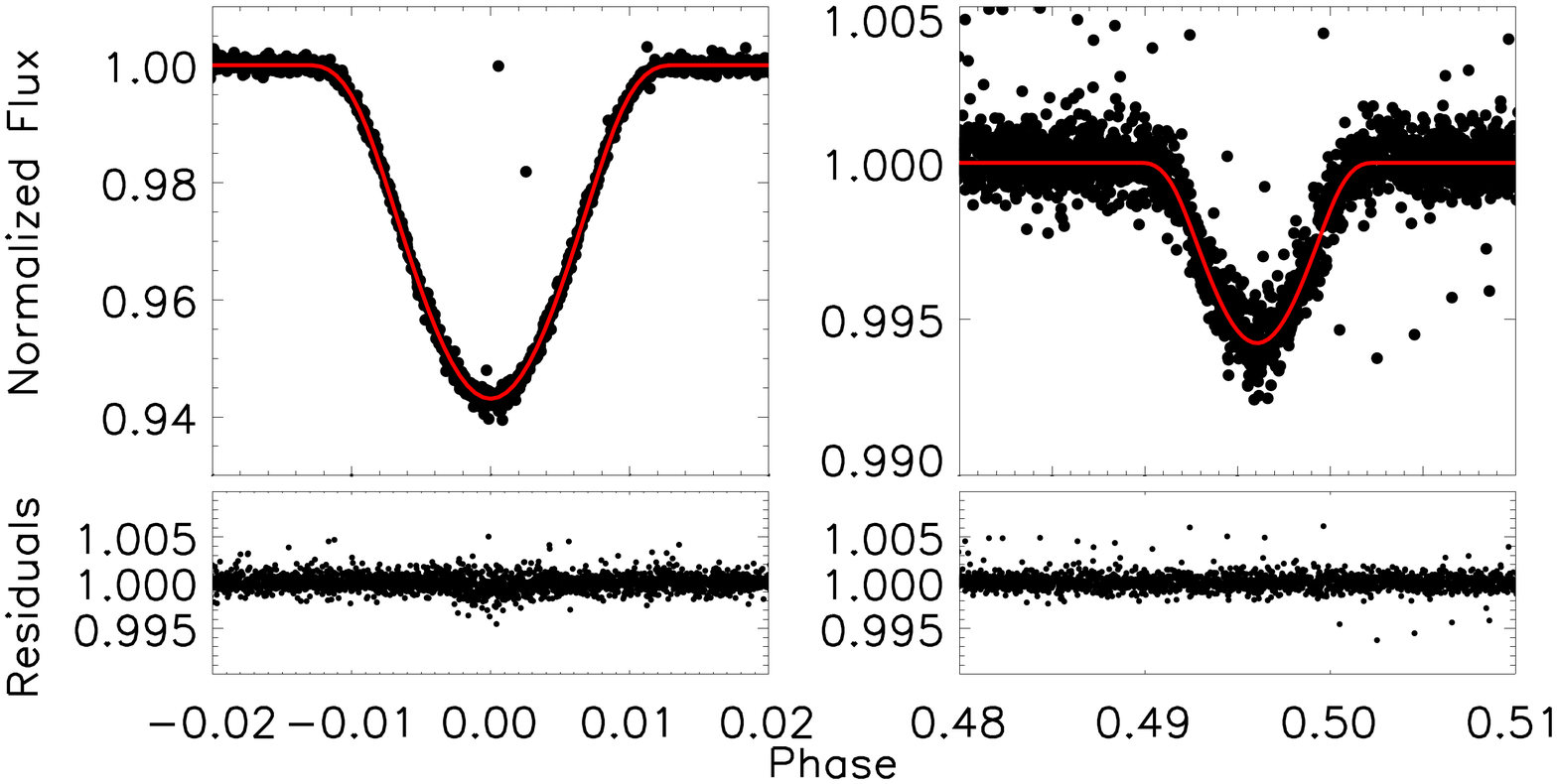}
\caption{
ELC model fits (solid lines) to the phase-folded primary (left) and secondary (right) stellar eclipses of \kic (black symbols); the lower panels show the best-fit residuals. The excess noise near the center of the transit is likely due to star spots.
\label{fig:ELC_lc}}
\end{figure}

Next, we phase-fold the light curve of \kic on our best-fit binary star period of $P = 10.116146$ days. For fitting purposes, we allow the limb-darkening coefficients of the primary star to vary freely. We note that star B is not completely occulted during the secondary stellar eclipse, and it's contribution to the total light during secondary eclipse needs to be taken into account. The best-fit models to the folded primary and secondary eclipses, based on the fast analytic mode of ELC (using \citealt{mand02}) are shown in Figure \ref{fig:ELC_lc}. The best-fit parameters for the ELC model of the {\em \kepler} light curve of \kic are listed in Table \ref{tab_parameters}. Including a ``third-light'' contamination of 8\% due to the nearby star (see \citealt{Kostov2014b}), we obtain $k = R_B/R_A = 0.5832\pm0.0695$, $a/R_A = 27.5438\pm0.0003$, $i_b = 87.3258\arcdeg\pm0.0987$\arcdeg, and $T_B/T_A = 0.7369\pm0.0153$. 

\begin{figure}
\centering
\plotone{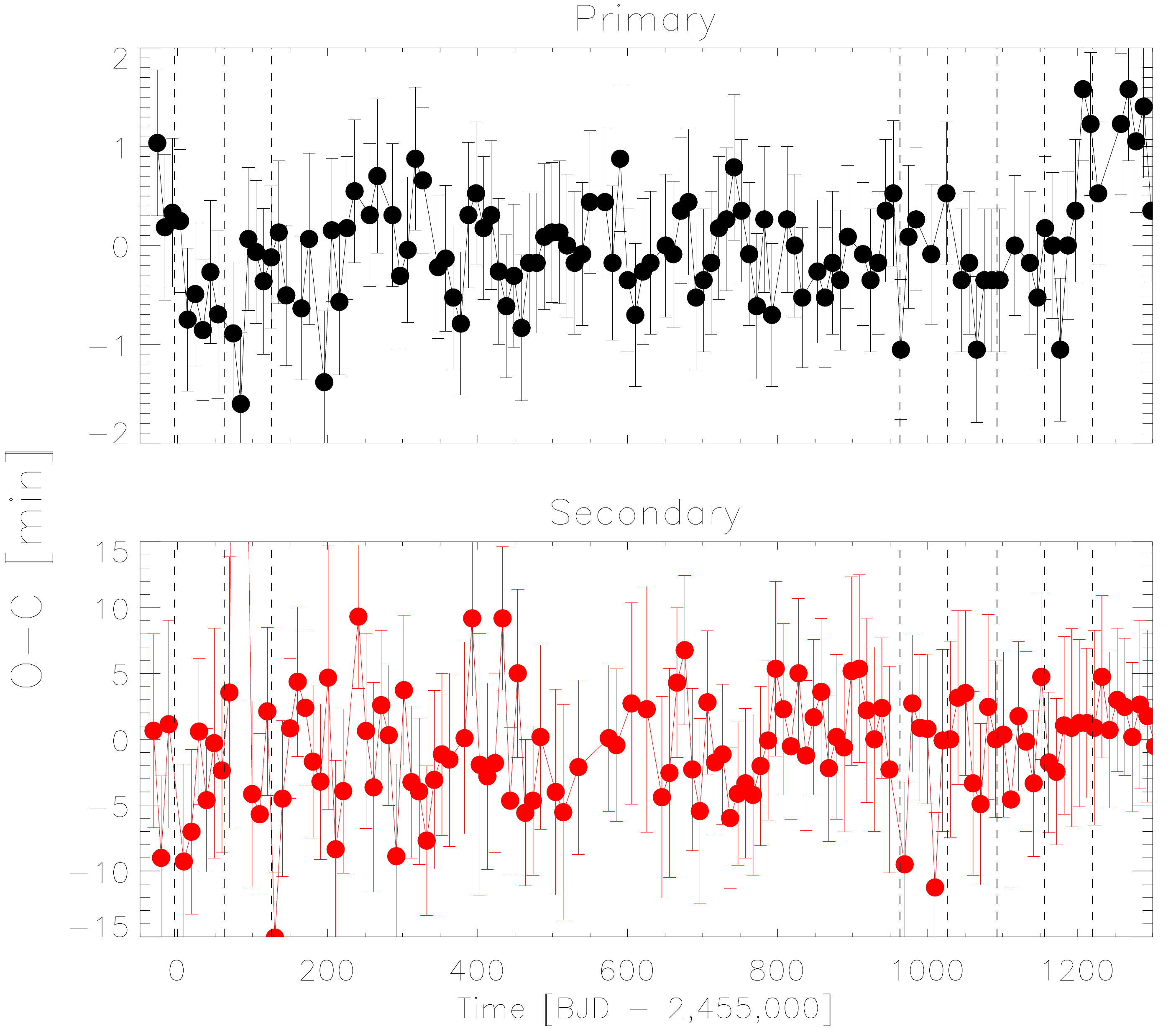}
\caption{Eclipse timing variations of the primary (upper panel, black) and secondary (lower panel, red, or grey color) eclipses of \kic, in terms of observed versus calculated (``O-C") deviations from linear ephemeris in minutes. The vertical dashed lines indicate the times of the planetary transits. The ``O-C'' deviations are consistent with noise, there are no discernible trends or periodicities. The last 8 points of the primary ETVs are excluded from our ETV analysis, as their anomalous shift by $\sim1$ min after day BJD 1200 is caused by a known\protect\footnotemark~absolute timing error of $\sim67$ sec for target pixel files from Quarters 1 through 14.
\label{fig:etvs}}
\end{figure}

We measure the stellar eclipse times using the methodology of \cite{Orosz2012}. For completeness, we briefly describe it here. We extract the data around each eclipse and detrend the light curve. Starting with the ephemeris given by our best-fit model, we phase-fold the light curve on the given period. Thus folded, the data were next used to create an eclipse template based on a cubic Hermite polynomial. Next, adjusting only the center time of the eclipse template, we iteratively fit it to each individual eclipse and measure the mid-eclipse times. To calculate eclipse time variations (ETVs), we fit a linear ephemeris to the measured primary and secondary eclipse times. The Observed minus Calculated ({\em ``O-C''}) residuals, shown in Figure \ref{fig:etvs}, have r.m.s. amplitudes of $A_{prim}\sim0.57$ min and $A_{sec}\sim8.6$ min respectively. Primary eclipses near days (BJD-2455000) 63, 155, 185, 246, 276, 337, 559, 640, 802, 842, 903, 994, 1015, 1035, 1105, 1126, 1237, and 1247 have been removed due to bad (with a flag of SAPQUALITY$\ne0$) or missing data. $A_{sec}$ is much larger than $A_{prim}$ because the secondary eclipses are much shallower than the primary eclipses and therefore is much noisier.

The high precision of the measured primary ETVs allow us to constrain the mass of the CBP. The planet contributes to non-zero ETVs through the geometric light travel-time and the dynamical perturbations it exerts on the EB \citep{bork13}. A CBP of $10M_{Jup}$ and with the orbital configuration of \kic would cause primary ETVs with amplitudes of $A_{geometric}\sim1.2$ sec and $A_{dynamic}\sim2.7$ min respectively. The latter is $\sim3\sigma$ larger than the measured amplitude of the primary ETVs, indicating an upper limit for the mass of the CBP of $\sim10M_{Jup}$, and thereby confirming its planetary nature. 

\footnotetext{$http://archive.stsci.edu/kepler/release\_notes/release\_notes21/DataRelease\_21\_20130508.pdf$}

\subsection{Discovering the transits of \kicb}

We discovered the planetary transits of \kicb using the method described in \cite{kostov2013}. For completeness, we briefly outline it here. 

Due to the aperiodic nature of the transits of a CBP, traditional methods used to search for periodic signals are not adequate. The amplitude of the transit timing variations between consecutive transits of \kicb, for example, are up to two days ($\sim3\%$ of one orbital period) compared to an average transit duration of less than 0.5 days. To account for this, we developed an algorithm tailored for finding individual box-shaped features in a light curve \citep{kostov2013}, based on the widely-used Box-fitting Least-Squares (BLS) method \citep{bls}. To distinguish between systematic effects and genuine transits, we incorporated the methodology of \cite{burke2006}.

Our procedure is as follows. Each detrended light curve is segmented into smaller sections of equal lengths (dependent on the period of the EB and on the quality of the detrending). Next, each section is replicated N times (the number is arbitrary) to create a periodic photometric time-series. We apply BLS to each and search for the most significant positive (transit) and negative (anti-transit, in the inverted time-series flux) box-shaped features. We compare the goodness-of-fit of the two in terms of the $\Delta_{\chi^{2}}$ difference between the box-shaped model and a straight line model. Systematic effects (positive or negative) present in a particular segment will have similar values for $\Delta_{(\chi^{2}){transit}}$ and $\Delta_{(\chi^{2}){anti-transit}}$. On the contrary, a segment with a dominant transit (or anti-transit) feature will be clearly separated from the rest on a $\Delta_{(\chi^{2}){transit}}$ versus $\Delta_{(\chi^{2}){anti-transit}}$ diagram. 

\begin{figure}
\centering
\plotone{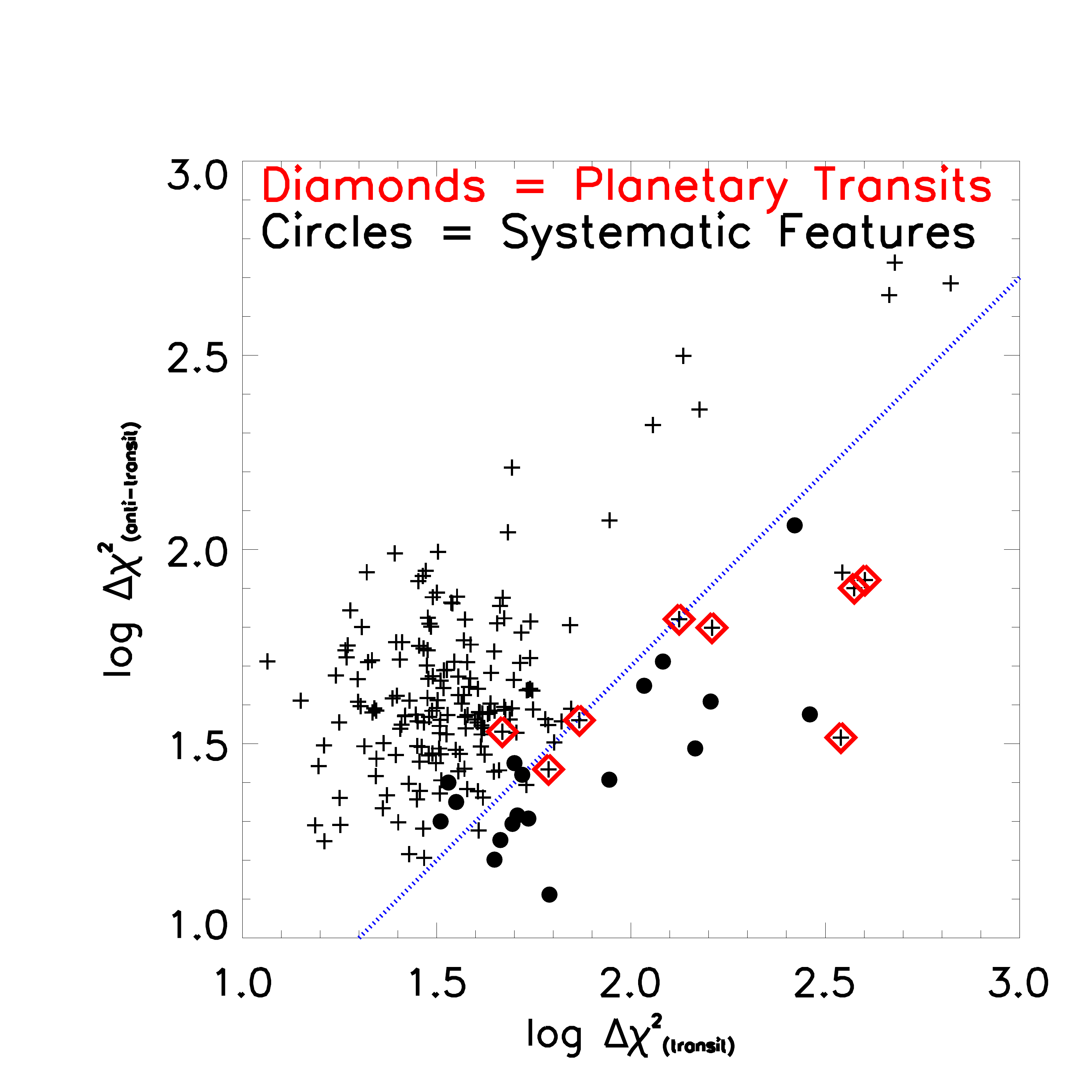}
\caption{
The (transit) -- (anti-transit) diagram for \kic. Each symbol represents the logarithmic ratio between the best transit and anti-transit signals detected in individual segments. The planetary transits are marked as red (or grey color) diamonds and the merit criterion line -- in blue. Black circles indicate segments where known systemic features mimic transits.
\label{fig:dipblip}}
\end{figure}

The (transit) -- (anti-transit) diagram for \kic is shown on Fig. \ref{fig:dipblip}. The segments of the light curve where no preferable signal (transit or anti-transit) is detected form a well-defined cloud, symmetrically distributed along the $\frac{\Delta_{(\chi^{2}){transit}}}{\Delta_{(\chi^{2}){anti-transit}}}=1$ line. The segments containing the transits of the CBP marked in red (or grey color) diamonds, along with a few other segments where systematic features dominate (black circles), exhibit a preferred $\Delta_{(\chi^{2}){transit}}$ signal. The blue line represents the merit criterion adopted for this target, defined in terms of an iteratively chosen ratio of $\frac{\Delta_{(\chi^{2}){transit}}}{\Delta_{(\chi^{2}){anti-transit}}}=2$. 

The signal for all but one (transit 7) of the \kicb transits is very strong. That transit 7 falls short of the criterion is not surprising. This event is the shortest and also the shallowest and can be easily missed even when scrutinized by eye. For \kic we had a preliminary dynamical model of the system based on events 1 through 6, prior to the release of Quarter 14 data. The observed events 7 and 8 were very near the predicted times, providing additional constraints to our model. 

\subsection{Stellar Rotation}
\label{sec:rotation}

Flux modulations of up to $\sim1\%$ on a timescale of $\sim13$ days are apparent in the light curve of \kic. We assume the source of this variation is star spots carried along with the rotation of the stellar surface of the primary, the dominant flux contributor ($\sim85\%$) in the \kepler~bandpass. To calculate the rotation period of star A, we compute Lomb-Scargle (L-S) periodograms and perform wavelet analysis \citep[using a Morlet wavelet of order 6,][]{torr98} for each Quarter separately. No single period matches all Quarters because of spot evolution as spots emerge/disappear (the most dramatic change, for example, being during Quarter 10). We estimate an average rotation period across all Quarters of $P_{rot,A}=13.1\pm0.3$ days and $P_{rot,A}=12.9\pm0.4$ days from Lomb-Scargle and wavelet analysis respectively. 

\begin{figure}
\centering
\plotone{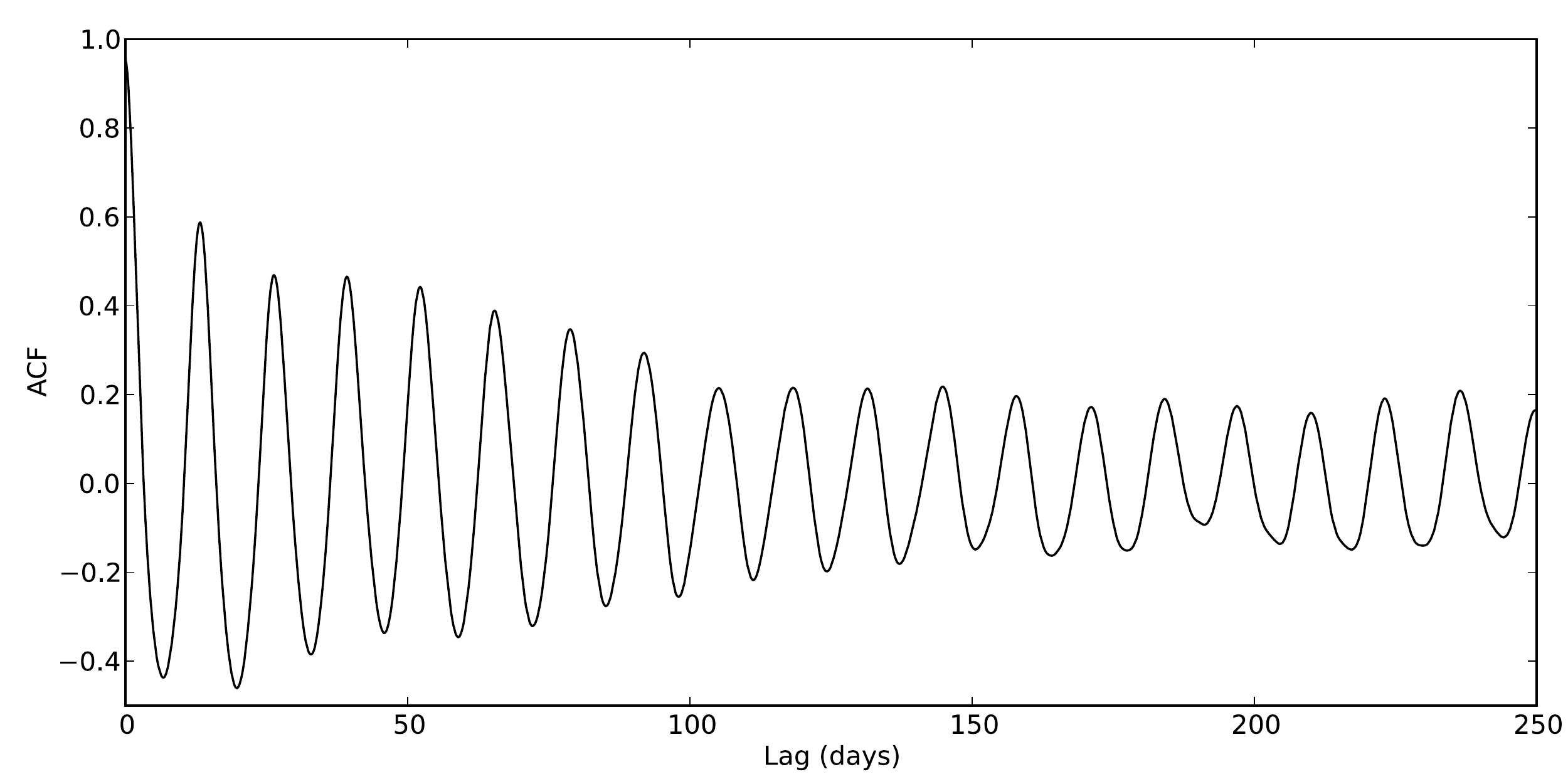}
\plotone{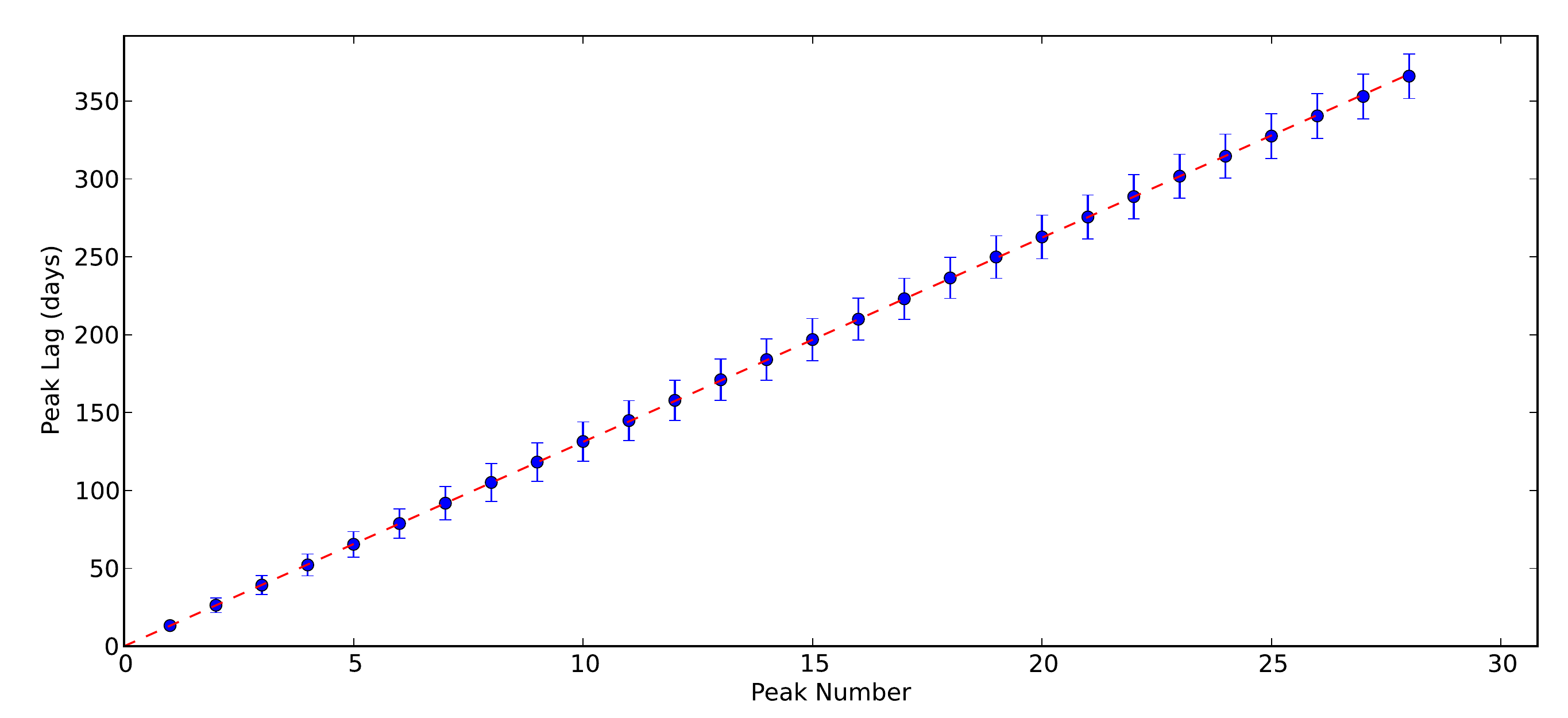}
\caption{
Upper panel: The autocorrelation function (ACF) of the cleaned and detrended light curve after the removal of the stellar eclipses. Lower panel: The measured lag of the ACF peaks (solid symbols), fitted with a straight line (dashed line). 
\label{fig:acf}}
\end{figure}

In addition, we measured the degree of self-similarity of the light curve over a range of different time lags by performing an autocorrelation function (ACF) analysis. In the case of rotational modulation, repeated spot-crossing signatures produce ACF peaks at lags corresponding to the rotation period and its integer multiples (McQuillan, Aigrain and Mazeh 2013). Figure \ref{fig:acf} depicts the autocorrelation function (ACF) of the cleaned and detrended light curve, after the primary and secondary eclipses were removed and replaced by the value of the mean light curve with a typical random noise. The autocorrelation reveals clear stable modulation with a period of about 13 days. To obtain a more precise value of the stellar rotation we measured the lags of the first 25 peaks of the autocorrelation and fitted them with a straight line, as shown in the lower panel of Figure \ref{fig:acf} (McQuillan, Mazeh and Aigrain 2013). From the slope of the fitted line we derived a value of $P_{rot,A}=13.15\pm0.15$ days as our best value for the stellar rotation period, consistent with the rotation period derived from the L-S analysis. 

We carefully inspected the light curve to verify the period and to ensure that it did not correspond to any harmonic of the spin period. A 13.1-day period matches the spot modulation well. Using the stellar rotation velocity measured from our spectral analysis we derive an upper limit to star A's radius of $R_A\le1.29~R_{\odot}$. The surface gravity of star A, $\Logg_A = 4.67$, provided by the NASA Exoplanet Archive\footnote{http://exoplanetarchive.ipac.caltech.edu}, in combination with the upper limit on $R_A$ indicate $M_A\le2.82~M_{\odot}$.

\subsection{Doppler Beaming}
\label{sec:doppler}

A radiation source emitting isotropically and moving at nonrelativistic speed with respect to the observer is subject to a Doppler beaming effect \citep{ryb79}. The apparent brightness of the source increases or decreases as it moves towards or away from the observer. To calculate the Doppler beaming factor for star A, we approximate its spectrum as that of a blackbody with $T_{eff} =4700$K (see next Section) and the \kepler~data as monochromatic observations centered at $\lambda=600$nm. Using Equations 2 and 3 from \cite{loeb2003}, we estimate the boost factor $3-\alpha = 5.13$. For the value of $K_1 = 43.49$ km s$^{-1}$ derived from the radial velocity, we expect a Doppler beaming effect due to star A with an amplitude of $\sim750$ ppm, on par with the intrinsic r.m.s. of the individual \kepler~measurements. The Doppler beaming contribution due to star B is much smaller (amplitude of $\sim50$ ppm) because of its small contribution to the total brightness of the system.

To search for the signal due to star A, we do a custom data detrending of the \kepler~light curve tailored to the rotational modulations. To each data point $t_i$, we fit either one or more sine waves with the same mean and period (but different phases and amplitudes) centered on the $[-0.5P_{rot,A}+t_i,t_i+0.5P_{rot,A}]$ interval. The mean values of the best-fit sine waves at each point represent a rotation-free light curve. Few sections of the light curve are consistent with a single spot (or group of) rotating in and out of view, and can be modeled with one sinusoid; most need two, or more. The continuously evolving spot pattern, the faintness of the source and the fact that the binary period is close to the rotation period of the primary star make detection of the otherwise strong expected signal ($\sim750$ ppm) challenging. Despite the custom detrending, the modulations in the processed data is consistent with noise and we could not detect the Doppler beaming oscillations caused by the motion of star A. We note that we successfully detected the Doppler beaming effect for Kepler-64 \citep{kostov2013}, where the amplitude is smaller but the target is brighter and the r.m.s. scatter per 30-min cadence smaller.
\section{Follow-up Observations}
\label{sec:followup}

\subsection{SOPHIE}
\label{SOPHIE}

\kic was observed in September-October 2012 and in March-April 2013 
with the SOPHIE spectrograph at the 1.93-m telescope of Haute-Provence 
Observatory, France. The goal was to detect the reflex motion of the primary star 
due to its secondary component through radial velocity variations. SOPHIE \citep{bouchy09} 
is a fiber-fed, cross-dispersed, environmentally stabilized 
\'echelle spectrograph dedicated to high-precision radial velocity measurements. 
The data were secured in High-Efficiency mode (resolution power $R=40\,000$) 
and slow read-out mode of the detector. The exposure times ranged between 1200 
and 1800~sec, allowing a signal-to-noise ratio per pixel in the range $4-8$ to be 
reached at 550~nm. The particularly low signal-to-noise ratio is due to the faintness 
of the target ($K_p = 15.52$mag).

\begin{figure}
\plotone{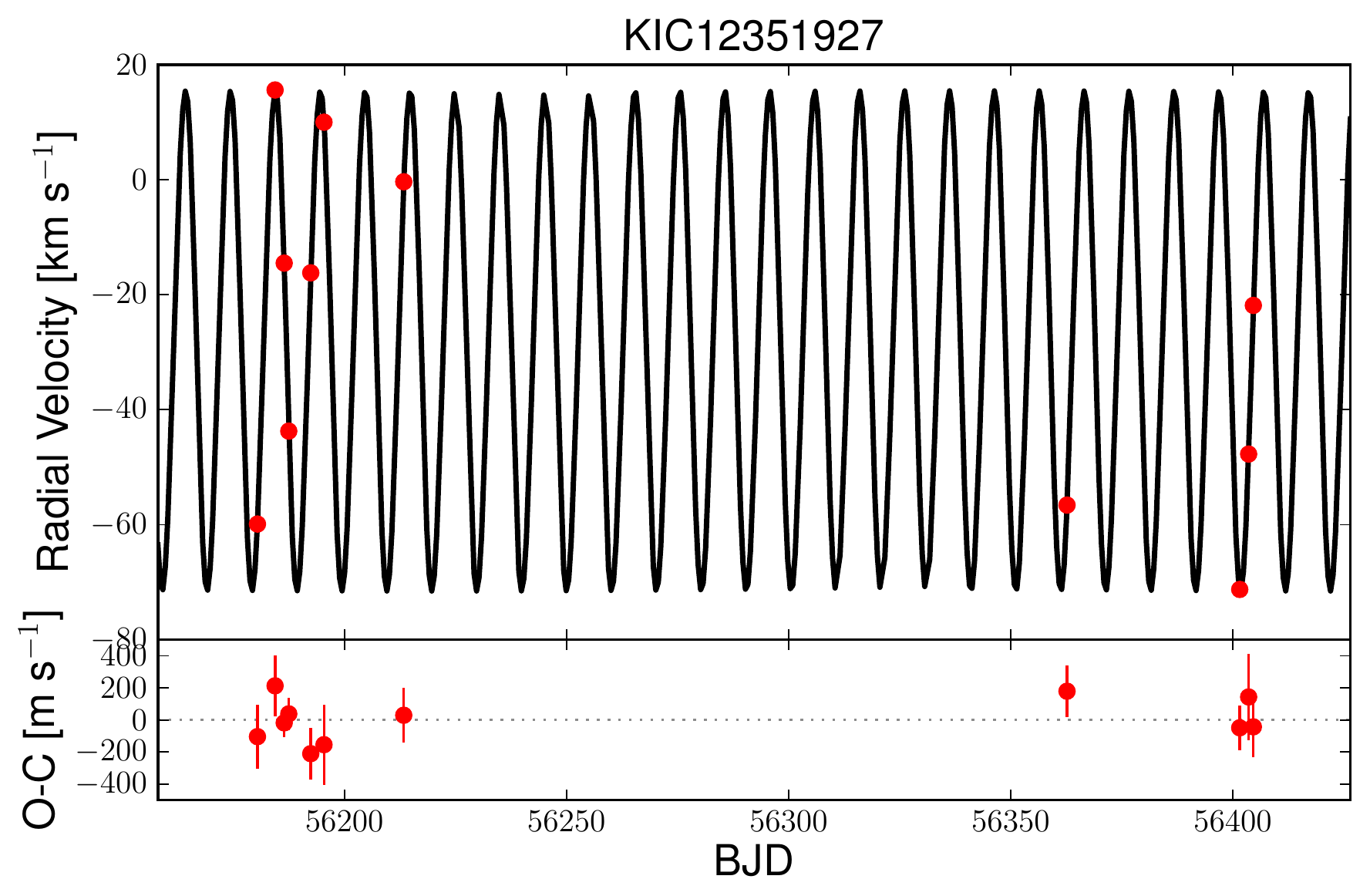}
\plotone{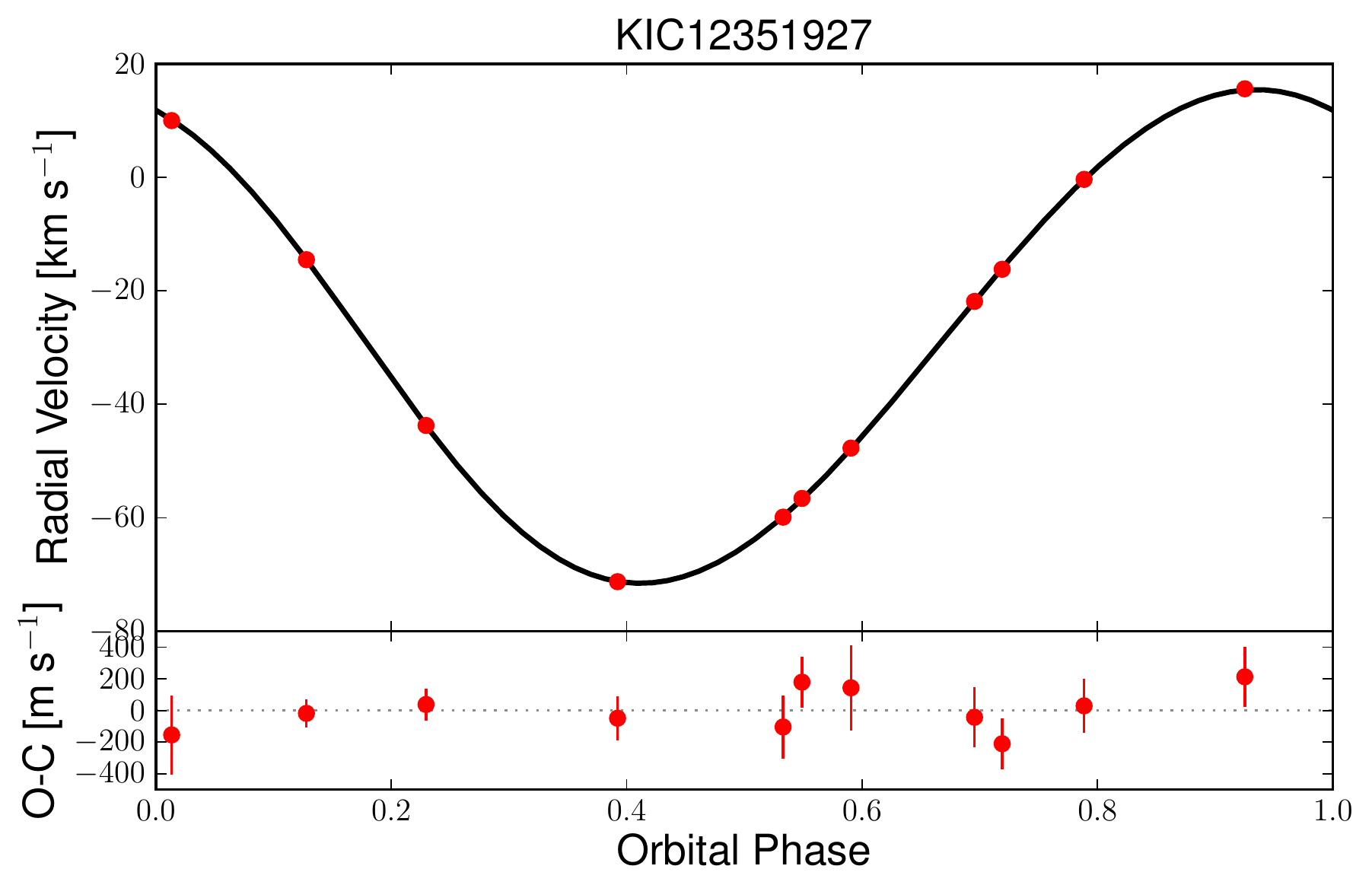}
\caption{SOPHIE radial velocity measurements of \kic with 1-$\sigma$\,Êerror bars as a function of time (upper) or orbital phase (lower) together with their Keplerian fit and residuals of the fit. Note the scale is km s$^{-1}$ for the radial velocities and m s$^{-1}$ for the O-C residuals.
\label{fig_orbits}
}
\end{figure}

The spectra were extracted from the detector images with the SOPHIE pipeline, which 
includes localization of the spectral orders on the 2D-images, optimal order extraction, 
cosmic-ray rejection, wavelength calibration and corrections of flat-field. Then we 
performed a cross-correlation of the extracted spectra with a G2-type numerical mask 
including more than 3500 lines. Significant cross-correlation functions (CCFs) were 
detected despite the low signal-to-noise ratio. Their Gaussian fits allow radial velocities 
to be measured as well as associated uncertainties, following the method described by 
\cite{baranne96} and \cite{pepe02}. The full width at half maximum (FWHM) of those 
Gaussians is $11 \pm 1$~km\,s$^{-1}$, and the 
contrast is $12 \pm 4$\,\%\  of the continuum. One of the observations 
(BJD$\,= 2\,456\,195.40345$) was corrected from the $230\pm30$\,m/s blue shift due 
to Moon light pollution and measured thanks to the reference fiber pointed on the sky 
\citep[e.g.][]{hebrard08}. The other exposures were not significantly 
polluted by sky background or by light from the Moon. The measured radial velocities are reported 
in Table~\ref{tab_RV} and plotted in Figure~\ref{fig_orbits}. Radial velocities show 
significant variations in phase with the \kepler~ephemeris. 

The radial velocities were fitted with a Keplerian model, taking into account the three 
constraints derived from the \kepler~photometry: the orbital period $P$, and the 
mid-times of the primary and secondary stellar eclipses, $T_{prim}$ and $T_{sec}$ respectively. 
The fits were made using the
\texttt{PASTIS} code \citep{diaz13}, previously used e.g. by \cite{santerne11} and \cite{hebrard13}. 
Confidence intervals around the best solutions were determined by Monte Carlo simulations. 
The histograms of the obtained parameters have a single-peak. 
We fitted them by Gaussians, whose centers and widths are the derived values and uncertainties reported in Table~\ref{tab_parameters}. The best fits are over-plotted with the data in Figure~\ref{fig_orbits}.
The dispersion of the residuals of the fit is 106\,m\,s$^{-1}$, in agreement with the error bars of the radial velocity measurements. We did not detect any significant drift of the radial velocities in addition to the reflex motion due to the binary. The small difference between the stellar eclipses, $T_{prim} - T_{sec}$, and $P/2$ measured from \kepler~photometry indicates that the orbit is not circular. Together with the radial velocities, it allows the detection of a small but significant 
eccentricity $e=0.037\pm0.002$, and longitude of the periastron $\omega=279.54\pm0.86\arcdeg$.
We note that our spectroscopic observations determined \kic as a single-lined spectroscopic binary, and allowed us to evaluate the binary mass function ${\it f(m)}$ from the derived radial velocity semi-amplitude of the primary star $K_1 = 43.485\pm0.085$~$km~s^{-1}$.

The signal-to-noise ratio of the final co-added spectrum is too low to allow a 
good spectral analysis of the star. The profile of the H-$\alpha$ line suggests an 
effective temperature $T_{\rm eff} \simeq 4700$~K. The width of the CCF implies 
$v \sin i_* =5\pm2\,$km\,s$^{-1}$.

\subsection{Third-light companion}
\label{photometry}

The large \kepler~pixel, $4\arcsec\times4\arcsec$ \citep{jen10b}, is prone to photometric contamination due to background sources. Unaccounted extra light inside the target's aperture can contribute to an erroneous interpretation of eclipse and transit depths, resulting in incorrect estimation of the relative sizes of the occulting objects. Proper characterization of such contamination is particularly important for the analysis of CBPs \citep[e.g.][]{Schwamb2013, kostov2013}. 

We note that there is a visible companion (``third light'') inside the central pixel of \kic at a separation of $\sim1.6\arcsec$ from the target, with a magnitude difference of $\Delta K_p\sim2.8$ \citep{Kostov2014b}. The presence of the companion can be deduced from 2MASS (Skrutskie et al. 2006) and UKIRT \citep{lawrence07} images, and from the full frame \kepler~image. A marked asymmetry in the target's point spread function, exhibited as a side bump with a position angle of $\sim218\arcdeg$, hints at the presence of an object close to \kic. 

During our reconnaissance spectroscopy with the 3.5-m Apache Point Observatory telescope we noticed the companion as a clearly separated star $\sim1.6\arcsec$ away from \kic. The companion was physically resolved using adaptive-optics-assisted photometry from Robo-AO \citep{baran13} and seeing-limited infrared photometry from WIYN/WHIRC \citep{meix10}. The measured flux contribution from the companion to the aperture of \kic is $\sim8\%, \sim15\%,\sim19\%$ and $\sim21\%$ in the {\em Kepler}, J-, H- and Ks-bands respectively \citep{Kostov2014b}; we correct for the contamination in our analysis. A detailed discussion of the companion's properties will be presented in future work \citep{Kostov2014b}.

The presence of such contamination is not unusual: adaptive-optics observations of 90 \kepler~planetary candidates show that $\sim20\%$ of them have one visual companion within $2\arcsec$ \citep{adams12}; lucky imaging observations by \cite{lillo12} find that $\sim17\%$ of 98 \kepler~Objects of Interest have at least one visual companion within $3\arcsec$. As more than 40\% of spectroscopic binaries with $P<10$ days are member of triple systems according to \cite{tok93}, it is reasonable to consider the visible companion to be gravitationally bound to \kic. Using Table 3 of \cite{gilli11}, for a contaminating star of $K_p\le18.5$~mag (i.e. $\Delta K_p\le3$~mag), and interpolating for the galactic latitude of \kic of $b=17.47\arcdeg$, we estimate the probability of a random alignment between a background source and \kic within an area of radius $1.6\arcsec$ to be $\sim0.002$. That despite the odds there is a star within this area indicates that the ``third light'' source is gravitationally bound to the EB, and could provide a natural mechanism for the observed misalignment of \kicb. Based on this statistical estimate, we argue that \kicb is a CBP in a triple stellar system. 
\section{Analysis of the system}
\label{sec:photodynamics}

A complete description of a CBP system requires 18 parameters -- three masses ($M_{A}$, $M_{B}$ and $M_p$), three radii ($R_{A}$, $R_{B}$, $R_P$), six orbital elements for the binary system ($a_{bin}, e_{bin}, \omega_{bin}, {\it i_{bin}}$, $\Omega_{bin}$ and phase $\phi_{0,bin}$ ) and six osculating orbital elements for the CBP ($a_p, e_p, \omega_p, {\it i_p}, \Omega_p$ and $\phi_{0,p}$). As described in Sections \ref{sec:kepler} and \ref{sec:followup}, some of these parameters can be evaluated from either the \kepler~data, or from follow-up photometric and spectroscopic observations. Measurements of the stellar radial velocities provide $e_{bin}, \omega_{bin}, {\it i_{bin}}$ and the binary mass function ${\it f(m)}$ (but not the individual stellar masses, as we observed \kic as a single-lined spectroscopic binary). The relative sizes of the two stars and the inclination of the binary system are derived from the \kepler~light curve. Based on the measured ETVs, we approximate the planet as a test particle ($M_p = 0$) for our preliminary solution of the system, and solve for its mass with the comprehensive photodynamical model. The value of $\Omega_{bin}$ is undetermined (see \citealt{doyle2011, Welsh2012}), unimportant to our analysis, and is set equal to zero. 

Here we derive the mass of the eclipsing binary (thus the masses of the primary and secondary stars) and the radius of the primary star from the planetary transits. Next, we produce a preliminary numerical solution of the system -- a necessary input for the comprehensive photometric dynamical analysis we present in Section \ref{sec:pd_model}. We study the dynamical stability of \kicb in Section \ref{sec:stability}.

\subsection{Initial Approach: Planetary transits and preliminary solutions}
\label{sec:pl_transits}

The mid-transit times, durations and depths of consecutive transits of a CBP are neither constant nor easy to predict when the number of observed events is low. However, while strictly periodic transit signals can be mimicked by a background contamination (either an EB or a planet), the variable behavior of CBP transits provide a unique signature without common false positives. 

Different outcomes can be observed depending on the phase of the binary system. While the CBP travels in one direction on the celestial sphere when at inferior conjunction, the projected velocities of each of the two stars can be in either direction. When the star and the planet move in the same direction, the duration of the transit will be longer than when the star is moving in the opposite direction with respect to the planet. As shown by \cite{kostov2013}, the transit durations as a function of binary phase can be used to constrain the it a priori unknown mass of the binary and the radius of the primary star (both critical parameters for the photodynamical model described below), assuming the planet transits across the same chord on the star. Typically, the more transits observed and the wider their EB phase coverage, the better the constraints are. 

While useful for favorable conditions, the approximation of \cite{kostov2013} is not applicable in general, and we extend it here. Depending on the relative positions of the CBP and the star on the sky, the CBP will transit across different chord lengths with associated impact parameters, such that different transits will have different durations and depths. A particular situation may be favorable, such as the cases of Kepler-64b and Kepler-47b where the CBPs transit across approximately constant chords. While the chords lengths do change from one transit to another, the variations are small as the stellar radii are sufficiently large, the mutual inclination between the orbits of the CBP and the EB is small, and the approximation in \cite{kostov2013} applies. The situation for \kic, however, is quite the opposite -- due to the misalignment between the two orbits and the small stellar radius, the chord length changes so much from one transit to another that the impact parameter is often larger than $R_{A}+R_{p}$, i.e. the planet misses a transit. To properly account for this novel behavior of a CBP, we modify our analytic approach accordingly to allow for variable impact parameter. Expanding on Equation (4) of \cite{kostov2013}, we add another term ({\it D}) to the numerator:

\begin{equation}
t_{dur,i} =  \frac{ABD_i}{1 + ACx_i}
\label{eq:durations}
\end{equation}

\begin{equation}
\begin{split}
A = (M_{bin})^{-1/3} \\
B = 2R_c (\frac{P_p}{2 \pi G })^{1/3} \\
C = - f(m)^{1/3} (\frac{P_p}{P_{bin}})^{1/3} (1-e^{2})^{-1/2} \\
D_i = \sqrt{1 - {\it b_i}^2} \\
x_i = (e\sin\omega + \sin(\theta_i + \omega))
\end{split}
\label{eq:durations2}
\end{equation}

{\noindent where $t_{dur,i}$, ${\it b_i}$ and $\theta_i$ are the duration, impact parameter and binary phase of the ${\it i_{th}}$ transit respectively, $M_{bin}$ is the sum of the masses of the two stars of the EB, $P_p$ is the average period of the CBP, $R_c = R_A + R_p$ is the transited chord length (where $R_A $ and $R_p$ are the radius of the primary star and the planet respectively), $f(m)$ is the binary mass function \citep[Eqn. 2.53,][]{hilditch01}, and {\it e} and $\omega$ are the binary eccentricity and argument of periastron respectively. Applying Equation \ref{eq:durations} to transits with {\it b\textgreater0} results in smaller derived $M_{bin}$ compared to transits across a maximum chord, {\it b=0}.}

The generally used method to derive {\it b} from the measured transit durations and depths for a planet orbiting a single star \citep{seager03} is not applicable for a CBP. The CBP impact parameter cannot be easily derived from the observables. From geometric considerations, {\it b} is:

\begin{equation}
b = \sqrt{(x_{s}-x_{p})^2 + (y_{s} - y_{p})^2}
\label{eq:b_CBP}
\end{equation}

{\noindent where ($x_{s}$, $y_{s}$) and ($x_{p}$,$y_{p}$) are the sky {\it x} and {\it y} - coordinates of the star and the planet respectively. The former depend on the binary parameters only and can be calculated from \cite{hilditch01}\footnote{Generally, $\Omega_{bin}$ (the EB longitude of ascending node) is undetermined and assumed to be zero}:}

\begin{equation}
\begin{split}
x_{s} = r_{s} \cos(\theta_{bin} + \omega_{bin}) \\
y_{s} = r_{s} \sin(\theta_{bin} + \omega_{bin}) \cos i_{bin}
\end{split} 
\label{eq:xy_EB}
\end{equation}

{\noindent where $r_{s}, \omega_{bin}, \theta_{bin}$ and $i_{bin}$ can be directly estimated from the radial velocity measurements and from the \kepler~light curve. The CBP coordinates, however, depend on the unknown mass of the binary and on the instantaneous orbital elements of the CBP $\Omega_{p}, \theta_{p}$ and $i_{p}$. Assuming a circular orbit for the CBP:}

\begin{equation}
\begin{split}
x_{p}=a_{p} [cos(\Omega_{p})cos(\theta_{p}) - sin(\Omega_{p})sin(\theta_{p})cos(i_{p})] \\
y_{p}=a_{p} [sin(\Omega_{p})cos(\theta_{p}) + cos(\Omega_{p})sin(\theta_{p})cos(i_{p})]
\end{split}
\label{eq:xy_CBP}
\end{equation}

{\noindent where $a_{p}$ is the semi-major axis of the CBP. For a mis-aligned CBP like \kicb, however, $\Omega_{p} \ne 0.0$ and equations \ref{eq:xy_CBP} cannot be simplified any further. In addition, due to 3-body dynamics, all three CBP orbital parameters vary with time. As a result, incorporating Equation \ref{eq:b_CBP} into Equation \ref{eq:durations} will significantly complicate the solution.}

However, we note that Equation \ref{eq:durations} uses only part of the information contained in the \kepler~light curve, i.e. transit durations and centers; it does not capitalize on the depth or shape of each transit. To fully exploit the available data, we evaluate the impact parameters of the eight transits directly from the light curve by fitting a limb-darkened transit model \cite{mand02} to each transit individually. The procedure is as follows. First, we scale the CB system to a reference frame of a mock, stationary primary star with a mass equal to the total binary mass of \kic. The scaling is done by adjusting for the relative velocities of the primary star \kic A ($V_{x,A}$), and of the CBP ($V_{x,p}$). The impact parameters are not modified by the scaling, as it does not change the distance between the planet and the star or their mutual inclination during each transit. We approximate $V_{x,p}$ as a single value for all transits:

\begin{equation}
V_{x,p} = ({\frac{2\pi GM_{bin}}{P_{p}}})^{1/3} = {\it const}
\label{eq:circ_vel}
\end{equation}
 
{\noindent A mock planet orbits the star on a circular, $P_{p}=66$ day orbit (the period of \kicb). The relative velocity of the observed CBP at the time of each transit $(V_{x,obs,i})$ is calculated as the absolute difference between the instantaneous $V_{x,p}$ and $V_{x,A}$:}

\begin{equation}
V_{x,obs,i} = |V_{x,p} - V_{x,A,i}|
\label{eq:scaled_vel}
\end{equation}

{\noindent where $V_{x,A,i}$ can be calculated from the fit to the RV measurements. The scaled time of the $i_{th}$ mock transit $t_{mock,i}$, referred to the time of minimum light, is then:}

\begin{equation}
t_{mock,i} = \frac{ |V_{x,p} - V_{x,A,i}|}{V_{x,p}} t_{obs,i}
\label{eq:scaled_time}
\end{equation}

{\noindent where $t_{obs,i}$ is the observed time during the $i_{th}$ transit. The mock transits are ``stretched'' with respect to the observed ones when $V_{x,A} < 0$ and ``compressed'' when $V_{x,A} > 0$. }

While $V_{x,p}$ depends on the unknown binary mass, it does so by only its third root (Equation \ref{eq:circ_vel}). For the low-mass binary we expect from the \kepler~Input Catalog, $V_{x,p}$ varies only by $\sim26\%$ for $M_{bin}$ between 1.0$M_{\odot}$ and 2.0$M_{\odot}$. Thus, the dominant factor in Eqn. \ref{eq:scaled_time} is $V_{x,A,i}$. 

The eight scaled, mock transits are next fit individually, sharing the same binary mass $M_{bin}$, size of the primary star $R_A$, and of the CBP radius $R_p$. The normalized semi-major axis of the mock planet, $a_{mock}/R_A$, depends on the binary phase of each transit and is different for different transits -- for fitting purposes it ranges from $(a_p - a_A)/R_A$ for transits near secondary stellar eclipse to $(a_p + a_A)/R_A$ for those near primary eclipse. Here $a_{p}$ is the mean semi-major axis of the CBP \kicb and $a_{A}$ is the semi-major axis of the primary star \kic -A. For light curve modeling, we use the limb-darkening coefficients from Section \ref{sec:kepler}. 

To estimate $R_{p}/R_{A}$, we first fit a limb-darkened light curve model to the scaled transit 8. The binary star is near a quadrature during the transit, $|V_{x,A,i}|$ is near zero, $a_{mock}\approx a_{p}$, $M_{bin}$ does not significantly affect Equation \ref{eq:scaled_time} and the scaling is minimal ($t_{mock,i}\approx t_{obs,i}$). To confirm that the scaling is negligible, we fit transit 8 for all $M_{bin}$ between 1.0 and 2.0. The differences between the derived values for $R_{p,8}/R_{A}$ are indistinguishable -- $R_{p,8}/R_{A} = 0.053$ for all $M_{bin}$, where $R_{p,8}$ is the radius of the planet deduced from the fit to scaled transit 8. We next use $R_{p,8}$ for light curve fitting of the other seven scaled transits. Also, the best-fit $a_{mock,8}$ from transit 8 is used in combination with $a_{A}$ to constrain the allowed range for $a_{mock,1-7}$ for the other seven transits, as described above. We note that while transit 1 also occurs near quadrature, the transit duration and depth are both much smaller than than those of transit 8, making the latter a better scaling ruler. The derived impact parameters for transits 1 through 8 are $0.85, 0.71, 0.17, 0.61, 0.84, 0.67, 0.78$ and $0.05$ respectively. We note that these are used to estimate $M_{bin}$ in Equation \ref{eq:durations} only and not as exact inputs to the photodynamical analysis described below.

\begin{figure}
\centering
\plotone{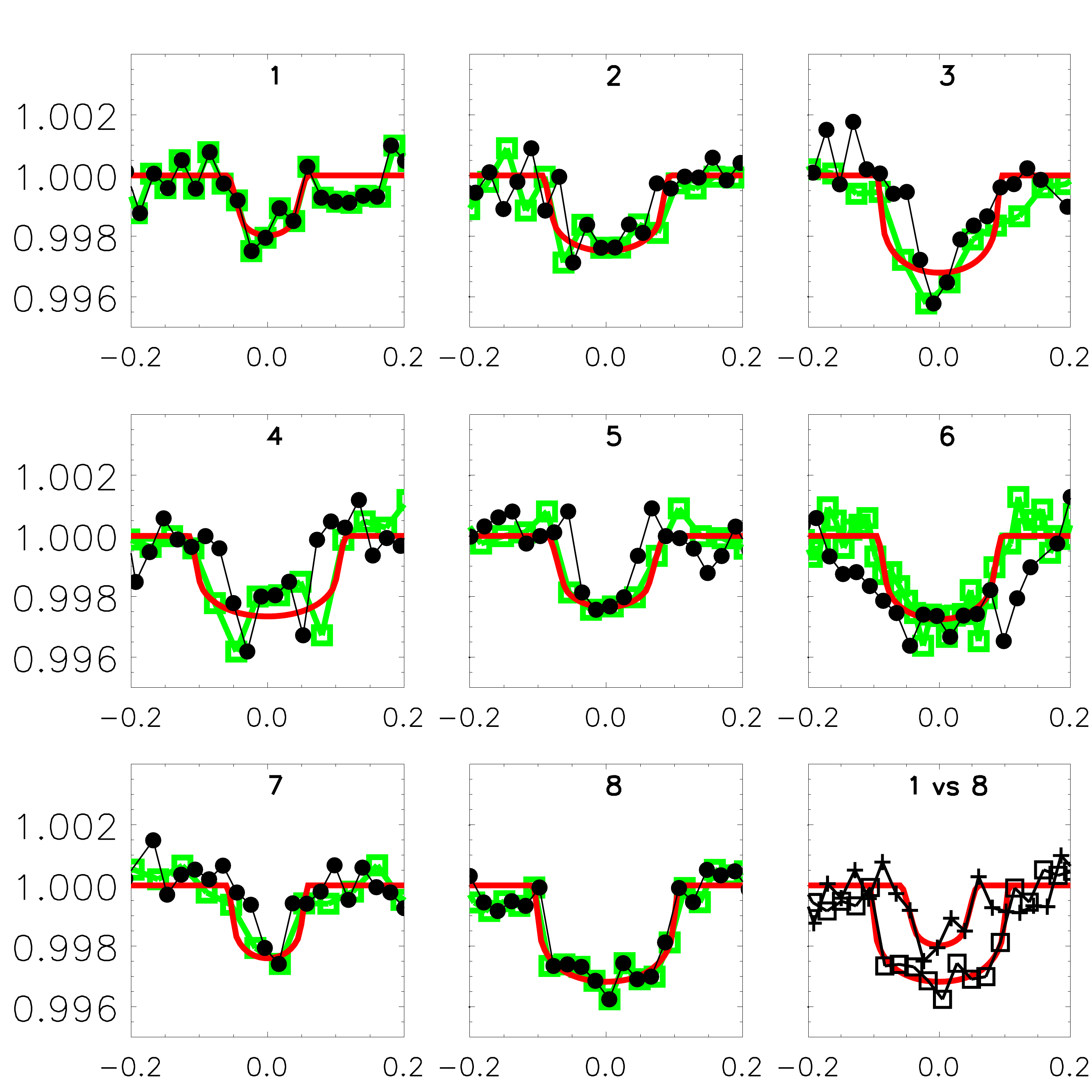}
\caption{Quadratic limb-darkened light curve model fits to the eight scaled transits of \kicb. Black symbols represents observed data, green (or light color) square symbols -- scaled data according to Eqn. \ref{eq:scaled_time} and red (or grey color) curve -- model fit to the scaled transits. We use the last transit (number 8) as a template for light curve fitting to estimate $R_{p}/R_{A}$ and $a_{p}$. The binary is near quadrature during transit 8, $V_{x,A}$ is at its lowest and the scaling used in Equation \ref{eq:scaled_time} is minimal. The result of orbital misalignment is represented in the last panel (``1 vs 8''), where we compare the two transits (square and cross symbols for Transits 8 and 1 respectively) that occur near the same EB phase, but have different impact parameters. }
\label{fig:scaled_transits}
\end{figure}

To evaluate the applicability of our approach, we test it on synthetic light curves designed to mimic  \kicb (8 transits, 10-11 misses, CBP on a $\sim66$-day orbit). For a noise-less light curve, we recover the simulated impact parameters of the 8 transits to within $0.01$, the semi-major axis to within 1\% and the size of the planet to within 10\%. Allowing the (known) mass of the simulated binary star to vary by $\pm0.5 M_{\odot}$ modifies the derived impact parameters by not more than 0.02. For a simulated set of light curves with normally distributed random noise of $\sim700$ ppm r.m.s. per 30-min cadence (similar to that of \kic) we recover the impact parameters to within $0.15$, the semi-major axis, and the size of the planet each to within 10\%. The good agreement between the derived and simulated model values validates the method. The observed (black) and scaled (green, or light color) transits of \kicb and the best-fit models (red, or grey color) to the latter are shown on Figure \ref{fig:scaled_transits}. 

We note that there are secondary effects not taken into account by Equation \ref{eq:scaled_time}. $V_{x,A}$, assumed to be constant in the equation, in reality varies throughout the duration of the transit. In principle, the longer the CBP transit, the more the stellar velocity and acceleration deviate from constancy. Longer transits (like Transit 6, see Figure \ref{fig:scaled_transits}) have asymmetric shape and the circular orbit approximation for the CBP in Equation \ref{eq:scaled_time} is not optimal. Depending on the phase of the binary star at the time of transit, both the magnitude and the sign of $V_{x,A}$ may change -- near quadrature, for example, the star changes direction.

\begin{figure}
\centering
\plotone{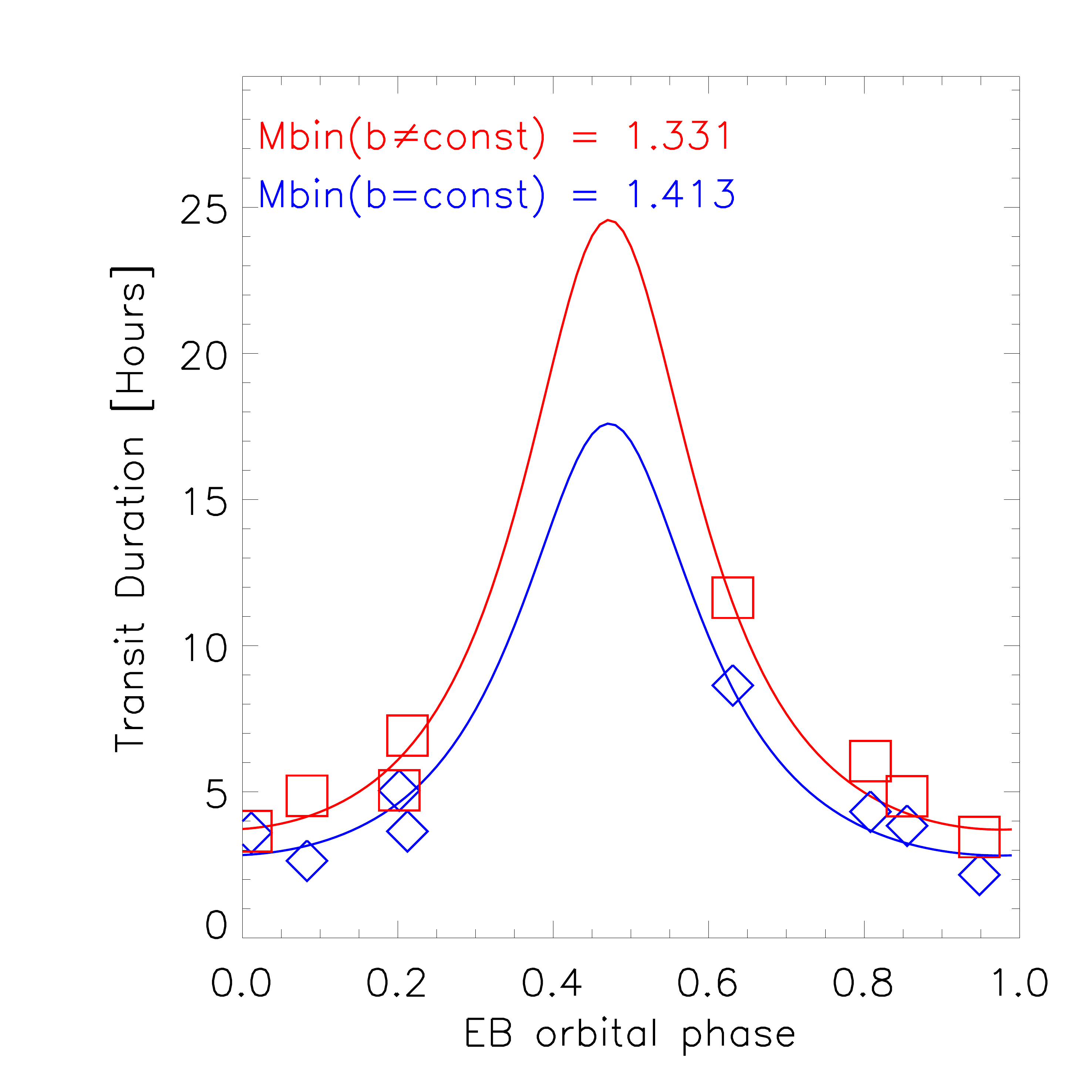}
\caption{CBP transit duration vs EB phase fits for \kicb using Equation \ref{eq:durations}. The blue and red curves represent the best fit for constant and for varying impact parameters respectively. The derived binary mass is 1$.41~M_{\odot}$ and $1.33~M_{\odot}$ for the two cases respectively. The derived primary radius is $0.70~R_{\odot}$ for the blue curve and $0.91~R_{\odot}$ for the red curve. Allowing for variable impact parameter results in a lower and higher estimates of $M_{bin}$ and $R_{A}$ respectively compared to the the case of constant impact parameter. }
\label{fig:dur_fit}
\end{figure}

Next, we apply Equation \ref{eq:durations} to the eight transits of \kicb for constant and for variable chords and compare the results. The best-fit models for the two cases are shown on Figure \ref{fig:dur_fit} as the blue and red curve respectively. The derived values for $M_{bin}$ and $R_{A}$ are $1.41~M_{\odot}$ and $0.70~R_{\odot}$ for constant {\it b} and $1.33~M_{\odot}$ and $0.91~R_{\odot}$ for varying {\it b}. Not accounting for different impact parameters overestimates $M_{bin}$ and underestimates $R_{A}$.

\begin{figure}
\centering
\plotone{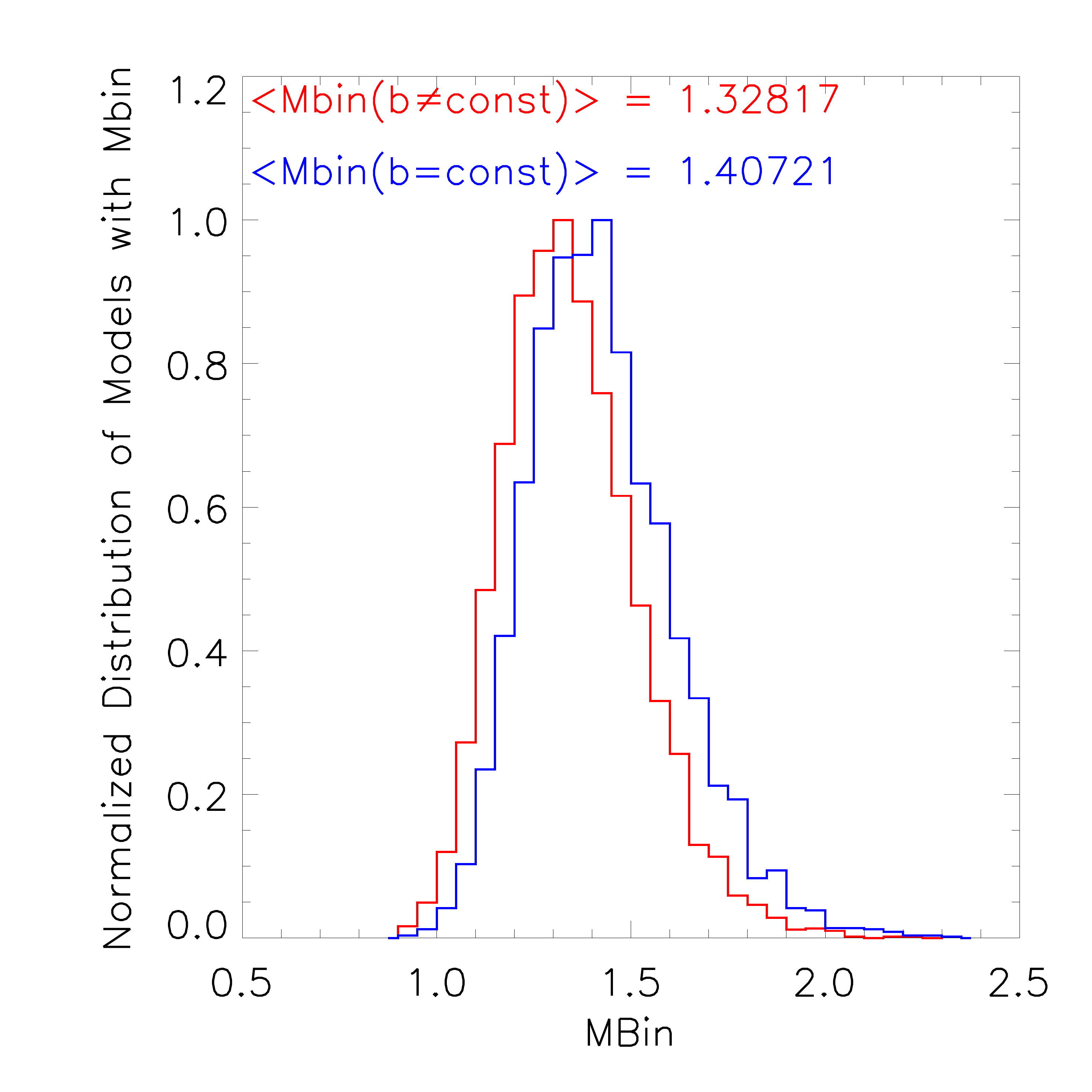}
\caption{Distribution of derived binary masses from Eqn. \ref{eq:durations} for a set of 10,000 scrambled observations. The blue histogram represents the distribution for constant impact parameters for all eight transits, and the red histogram -- for different impact parameters.}
\label{fig:scrambled_durations}
\end{figure}

We use the measured transit duration uncertainties to constrain the derived binary mass as follows. We simulate a set of 10,000 scrambled observations, each consisting of the eight measured transit durations individually perturbed by adding a normally distributed noise with a standard deviation of $20$ min. Next, we apply Equation \ref{eq:durations} to each realization. The distribution of the derived $M_{bin}$ for the entire set of scrambled observation is shown in Figure \ref{fig:scrambled_durations}. The blue histogram represents the solutions accounting for constant chord length and the red histogram -- for variable chord length. The median values for binary mass and their 1-sigma deviations are $1.41\pm0.19~M_{\odot}$ and $1.33\pm0.17~M_{\odot}$ for the former and latter case respectively. Based on these results, for our preliminary photodynamical search over the parameter space of the \kic system (described next) we adopt the latter case, and allow the binary mass to vary from 1.16 to 1.5 $M_{\odot}$.

For our initial photodynamical solutions we use a numerical N-body integrator \citep[described in][]{kostov2013} to solve the equations of motion. For completeness, we briefly outline it here and discuss the modifications we added for diagnosing \kic. The integrator is an implementation of the SWIFT code\footnote{http://www.boulder.swri.edu/~hal/swift.html} adapted for IDL. Due to the particular behavior of the CBP transits of \kicb, we can neither fix the planetary inclination $i_{p}$ to 90 degrees, or the ascending node $\Omega_p$ and the initial phase $\phi_{0,p}$ to zero. Unlike the case of Kepler-64b described in \cite{kostov2013}, here we solve numerically for these three parameters. Furthermore, it is not optimal to choose the time of the first transit as the starting point of the numerical integration as \cite{kostov2013} did. Doing so would introduce an additional parameter -- the impact parameter ${\it b_0}$ of the chosen transit; the estimated impact parameters of the individual transits indicated above are too coarse to be used in the photodynamical model. Instead, here we specify initial conditions with respect to the time when the planet is crossing the x-y plane ({$z_p=0$}), approximately 3/4 of a planetary period prior to transit 2, i.e. at $t_0 = 2,455,014.465430$ (BJD). This allows us to find the true anomaly of the planet ($\theta_p = 2 \pi - \omega_p$), and the planet's eccentric and mean anomalies at the reference time. The number of free parameters we solve for is 9: [$ M_{A}, a_p, e_p, \omega_p, {\it i_p}, \Omega_p, \phi_{0,p}, R_{A}$ and $R_{p}$].

Restricting the binary mass to the $1\sigma$ range indicated by the scrambled durations, we fit preliminary photodynamical models to the eight transits of \kicb by performing a grid search over the 8 parameters. The quality of the fit is defined as the chi-square value of the observed minus calculated (O-C) mid-transit times of all 8 events. Starting with an initial, coarse time step of 0.1 days, we select the models that reproduce the mid-transit times of each of the observed eight transits to within 0.05 days and also correctly ``miss'' all other events by more than $R_{A}+R_p$. Next, we refine the grid search by reducing the time step to 0.02 days, and minimize again. The best-fit model is further promoted for a detailed MCMC exploration as described in the next section. 

\subsection{Comprehensive photometric-dynamical analysis}
\label{sec:pd_model}

The {\em \kepler} light curve and radial velocity data for \kic were further modeled using a comprehensive photometric-dynamical model. This model uses a dynamical simulation, assuming only Newton's equations of motion and the finite speed of light, to predict the positions of the stars and planet at the observed times \citep[e.g.,][]{doyle2011,Welsh2012}. The parameters of this simulation are functions of the initial conditions and masses of the three bodies, and are provided by the preliminary simulations described above. These positions are used as inputs -- along with radii, limb darkening parameters, fluxes and ``third-light'' contamination -- to a code \citep{josh11,pal12} that produces the modeled total flux (appropriately integrated to the {\em Kepler} `long cadence' exposure). This flux is compared directly to a subset of the full {\em Kepler} data. The radial velocity data of the larger star are compared to the velocities determined by the dynamical simulation at the observed times.

We isolate only the {\em Kepler} data within a day of the stellar eclipses or suspected planetary transit crossing events (data involving `missing' events are included as well). Those data, excluding the eclipse features, are divided by a linear function in time in order to detrend the light curve for local astrophysical or systematic features that are unrelated to the eclipses.  

The model described in this section has 23 adjustable parameters. Three parameters are associated with the radial velocity data: the RV semi-amplitude of star A, $K_A$, the RV offset, $\gamma_A$, and a `jitter' term, $\sigma_{\rm RV}$, that is added in quadrature to the individual RV errors, correcting for unaccounted systematic error sources. The initial conditions are provided as instantaneous Keplerian elements of the stellar (subscript {\it ``bin''}) and planetary (subscript {\it ``p''}) orbits, defined in the Jacobian scheme: the periods, $P_{bin,p}$, the sky-plane inclinations $i_{bin,p}$, vectorial eccentricities $e_{bin,p} \cos(\omega_{bin,p})$, $e_{bin,p} \sin(\omega_{bin,p})$, the relative nodal longitude $\Delta \Omega = \Omega_p -\Omega_{bin}$ and the times of barycenter passage $T_{bin,p}$. The latter parameters are more precisely constrained by the data than the mean anomalies; however, they may be related to the mean anomalies, $\eta_{bin,p}$, via
\begin{eqnarray}
	\frac{P_{bin,p}}{2 \pi} \eta_{bin,p} &=& t_0 - T_{bin,p}+ \frac{P_{bin,p}}{2 \pi} \left[E_{bin,p}-e_{bin,p}\sin(E_{bin,p})\right]
\end{eqnarray}
where $E_{bin,p}$ are the eccentric anomalies at barycenter passage, defined by
\begin{eqnarray}
	\tan\left(\frac{E_{bin,p}}{2}\right) & = & \sqrt{\frac{1-e_{bin,p}}{1+e_{bin,p}}} \tan\left(\frac{\pi}{4}-\omega_{bin,p}\right)
\end{eqnarray}
Two parameters are the mass ratios between stars and planet, $M_A/M_B$ and $M_p/M_A$. The remaining 7 parameters are related to the photometric model: the density of star A, $\rho_A$, the two radii ratios, $R_B/R_A$ and $R_b/R_A$, the \kepler-band flux ratio $F_B/F_A$, the linear limb darkening parameter of star A, $u_1$, and the additional flux from contaminating sources $F_X/F_A$. A final parameter parameterizes the Gaussian distribution of the photometric residuals, $\sigma_{\rm LC}$.

We adopted uniform priors in all the parameters excluding the vectorial eccentricities and $F_X/F_A$. For those parameters we enforced uniform priors in $e_{bin,p}$ and $\omega_{1,2}$ and a Gaussian prior in $F_X/F_A$ with mean $0.08$ and variance $0.0001$. The likelihood of a given set of parameters was defined as
\begin{eqnarray}
	L & \propto & \prod^{N_{\rm LC}} \sigma_{\rm LC}^{-1} \exp\left[-\frac{\Delta F_i^2}{2 \sigma_{\rm LC}^2}\right] \\ \nonumber
		& & \times \prod^{N_{\rm RV}} \left(\sigma_{\rm RV}^2+\sigma_i^2\right)^{-1/2} \exp\left[-\frac{\Delta {\rm RV}_i^2}{2\left( \sigma_{i}^2+\sigma_{\rm RV}^2\right)}\right]
\end{eqnarray}
where $\Delta {\rm LC}_i$ is the residual of the $i$th photometric measurement and $\Delta {\rm RV}_i$ is the residual of the $i$th radial velocity measurement with formal error $\sigma_i$.

We explored the parameter space with a Differential Evolution Markov Chain Monte Carlo (DE-MCMC) algorithm \cite{terBraak08}. In detail, we generated a population of 60 chains and evolved through approximately 100,000 generations.  The initial parameter states of the 60 chains were randomly selected from an over-dispersed region in parameter space bounding the final posterior distribution. The first 10\% of the links in each individual Markov chain were clipped, and the resulting chains were concatenated to form a single Markov chain, after having confirmed that each chain had converged according to the standard criteria including the Gelman-Rubin convergence statistics and the observation of a long effective chain length in each parameter (as determined from the chain autocorrelation). 

\begin{figure}
\centering
\plotone{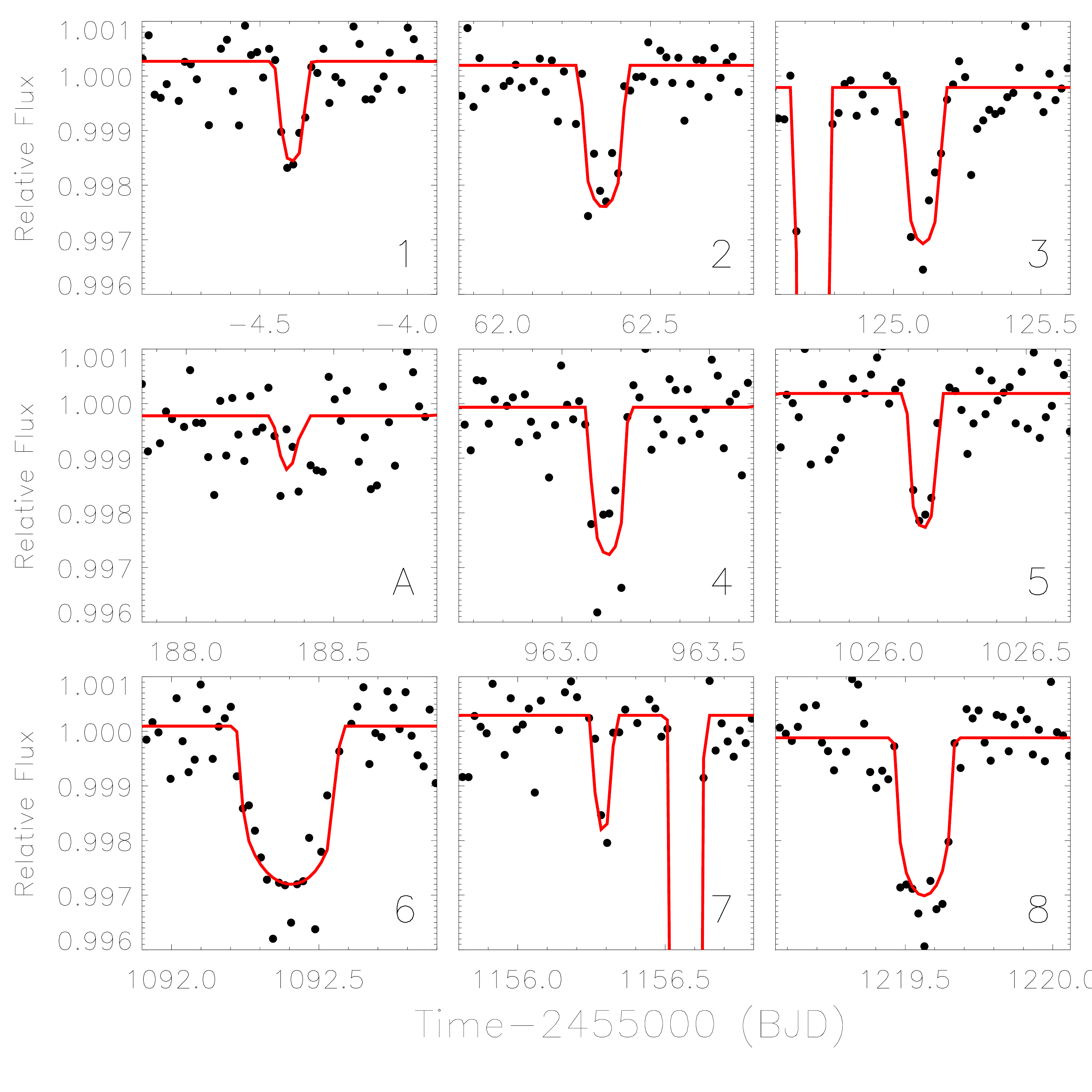}
\caption{
Photodynamical fits (red, or grey color) to the 8 observed (and to a predicted ninth, labeled as ``A'' near time 188.35 (BJD - 2,455,000), very shallow and buried in the noise) transits (black symbols) for the best-fit model in Tables \ref{tab:pd_in} and \ref{tab:pd_out}. Stellar eclipses are also shown at times 124.7 and 1156.5 (BJD - 2,455,000). We note the timescale between transits 3 and 4.
\label{fig:photodyn_hits}}
\end{figure}

\begin{figure}
\centering
\plotone{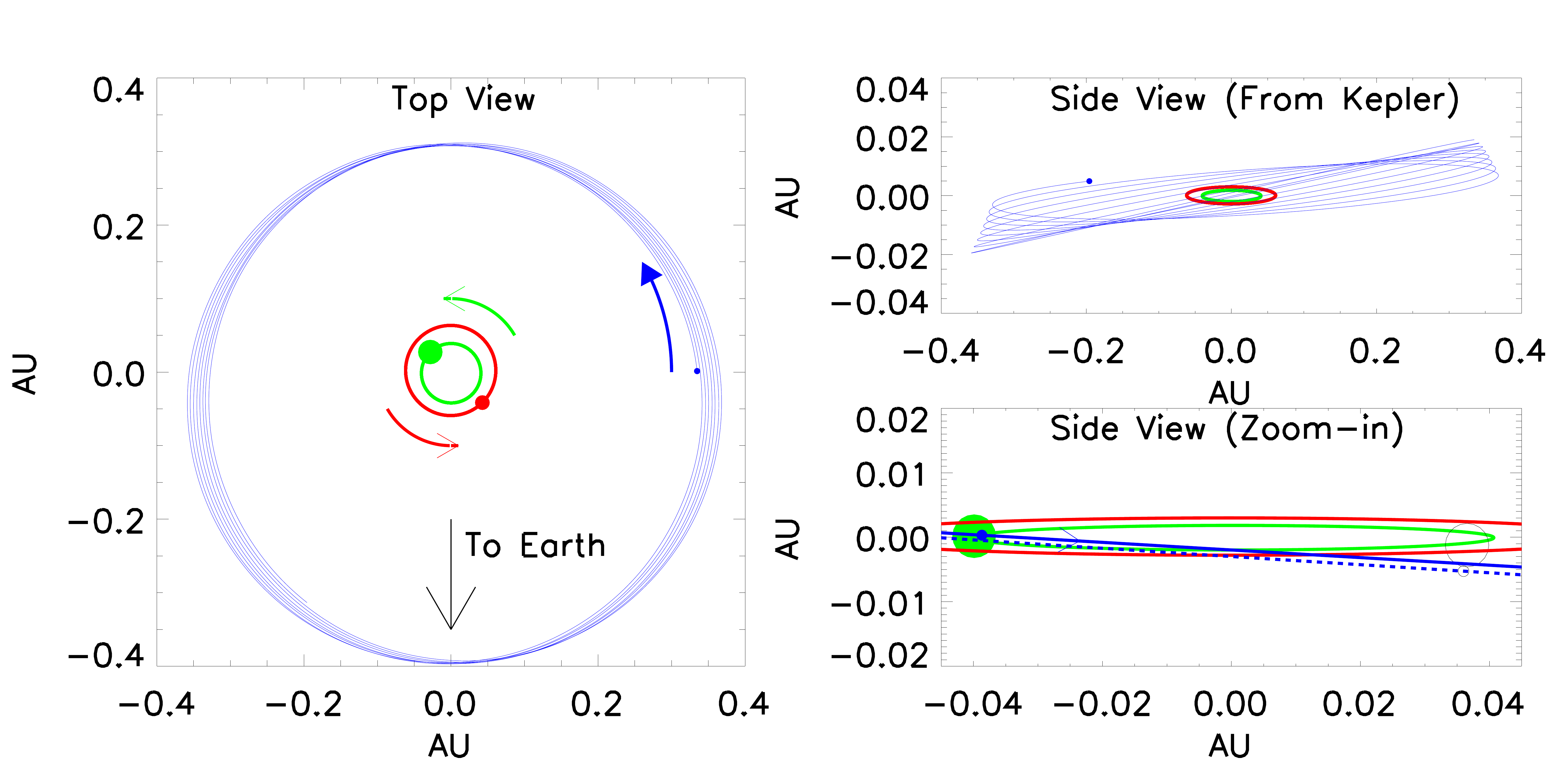}
\caption{Orbital configuration of \kicb over the course of $1/8$ precession period ($1/8$ of $\sim11$ years). The orbits of the primary (green, or light color) and secondary (red, or grey color) stars, and of the CBP (blue, or dark color), are to scale on the left and lower right panels. The EB symbols in the left panel, the CBP symbols and the vertical axis in the upper right panel are exaggerated by a factor of 5, 5, and 10 respectively. The EB symbols in the lower right panel are to scale. The precession of the argument of periastron of the CBP ($\omega_p$) as it increases by $90\arcdeg$ is clearly seen in the left panel. Two consecutive passages of the CBP at inferior conjunction are shown in the lower right panel, demonstrating a missed transit: 
the solid overlapping symbols (and blue, or dark color line for the sky path of the CBP) illustrate the configuration of the system at the last observed transit (transit 8) and, one planetary period later, one missed transit (open symbols for the primary star and the CBP respectively).}
\label{fig:123_orbit}
\end{figure}

The photodynamical fits to the 8 observed transits of the CBP are shown in Figure \ref{fig:photodyn_hits}. We note that our model predicts a ninth, very shallow and buried in the noise transit, labeled as ``A'' in Figure \ref{fig:photodyn_hits}. For clarity, we label the observed transits with a number, and those either missed or not detected with a letter. We tabulate the results of this analysis in Tables \ref{tab:pd_in} and \ref{tab:pd_out}, reporting the median and 68\% confidence interval for the marginalized distributions in the model parameters and some derived parameters of interest. The parameters we adopt throughout this paper are the ``best-fit'' values reported in Tables \ref{tab:pd_in} and \ref{tab:pd_out}. The orbital configuration of the system is shown on Figure \ref{fig:123_orbit}. The orbit of the CBP evolves continuously and, due to precession, is not closed. We note that our best-fit mass for the planet is large for it's radius. The expected mass is $M_p\sim16M_\oplus$, using the mass-radius relation of \cite{weiss13} for $1M_\oplus<M<150M_\oplus$, whereas our model provides $M_p\sim67M_\oplus\pm21M_\oplus$. This suggests that either \kicb is a much denser planet (mix of rock, metal, gas), or that the mass is even more uncertain than stated, and a factor of 2-3 times likely smaller.

We note that the binary orbit reacts to the gravitational perturbation of the planet. As a result, the EB orbital parameters and eclipse times are not constant. The effect, however, is difficult to measure with the available data. Also, the planetary orbit does not complete one full precession period between transits 1 and 8. The precession period for our best-fit model is $\sim4000$ days, in line with the analytic estimate of $\sim4300$ days (for equal mass stars) based on \cite{schneider1994} . After transit 8, the transits cease as the planetary orbit precesses away from the favorable transit configuration. The transits will reappear after BJD 2458999 (2020 May 29). 

\subsection{Orbital Stability}
\label{sec:stability}

The minimum critical semi-major axis (\cite{holman99}, Eq. 3) for the best-fit parameters of the \kic system is $a_{crit} = 2.55~a_{bin} = 0.26$~AU. With a semi-major axis that is $\approx37\%$ larger than the critical limit ($a_p = 0.3553$~AU), the orbit of the planet \kicb is in a gravitationally stable region. We note that due to the planet's non-zero eccentricity, its closest approach to the binary is reduced by $(1-e)$ and the stability criterion is more tight -- $r_{p,min} = a_p\times(1-e_p) = 0.3168$~AU, closer compared to a zero-eccentricity orbit but still beyond $a_{crit}$.

Three-body systems are notorious for exhibiting complex dynamical behavior spurred by mean-motion resonances (MMR). To explore the long-term stability of the \kic system we have studied its dynamical behavior by utilizing the MEGNO\footnote{Mean Exponential Growth of Nearby Orbits} factor $\langle Y\rangle$ \citep{cincottasimo2000a,cincottasimo2000b,cincottaetal2003}, a widely used method for dynamical analysis of mutliplanet systems \citep{Gozdziewski2008,Hinse2010}. We note that by a stable orbit here we refer to an orbit that is stable only up to the duration of the numerical integration, i.e. a quasi-periodic orbit. The time scale we use is sufficient to detect the most important mean-motion resonances. However, the dynamical behavior of the system past the last integration time-step is unknown. 

\begin{figure}
\centering
\plotone{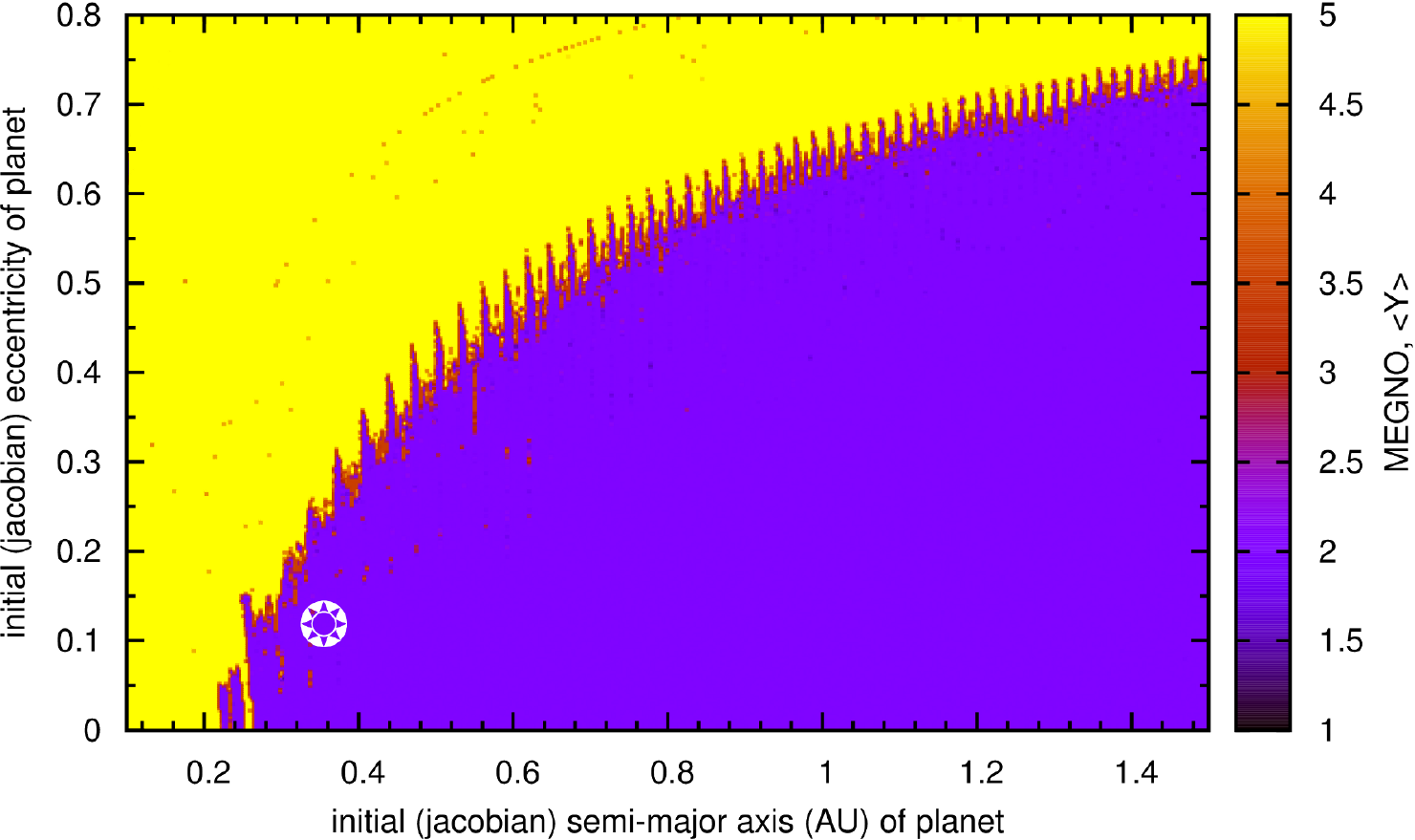}
\plotone{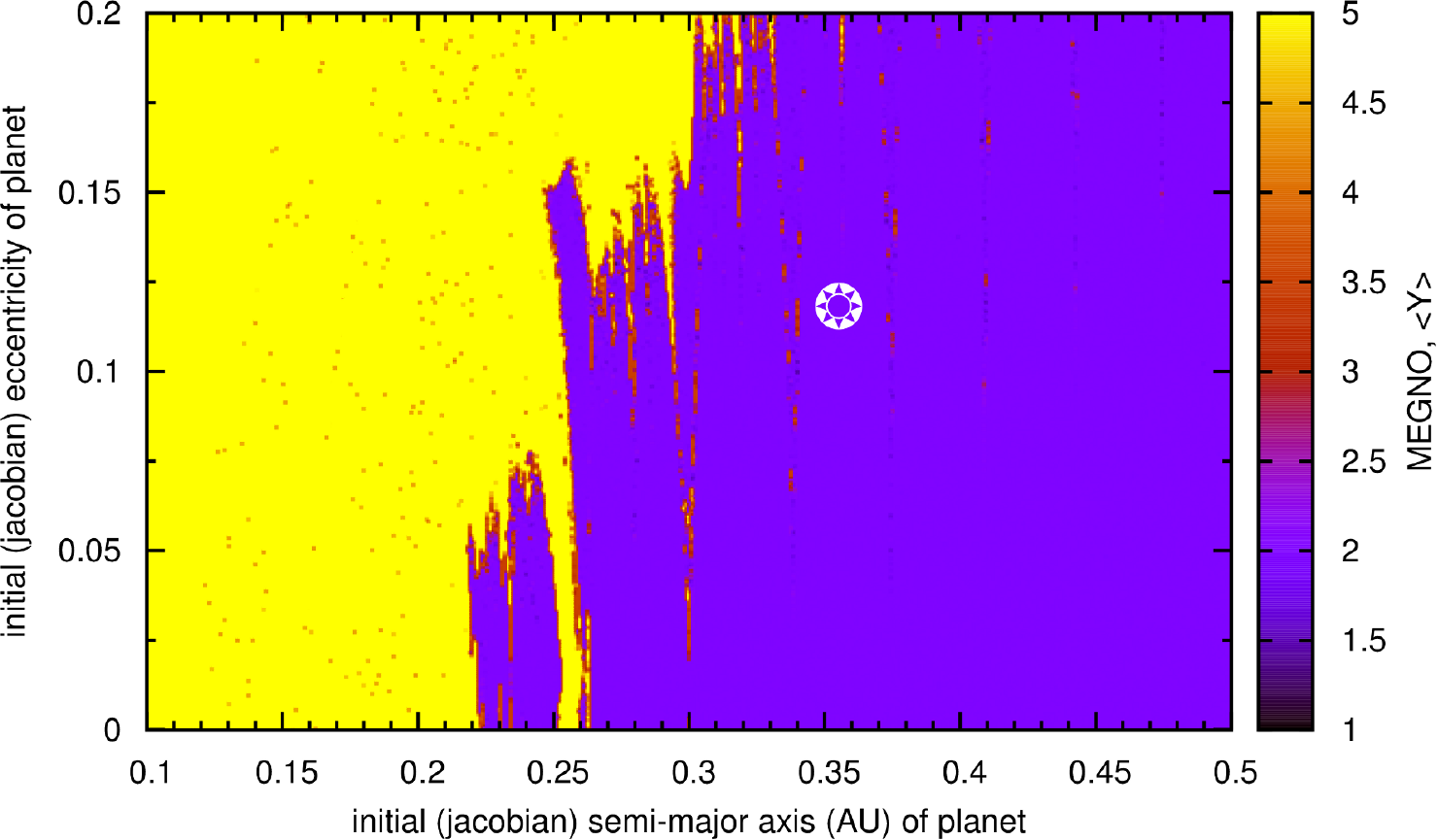}
\caption{
Upper panel: MEGNO map of \kicb, using the best-fit parameters from Table \ref{tab:pd_out}. 
Purple (or dark) color outlines quasi-periodic regions in the $(a,e)$-space, and yellow (or light) color -- chaotic (and possibly unstable) regions (see text for details). The cross-hair mark denotes the instantaneous Jacobian coordinates of the planet, placing it firmly in a quasi-periodic orbit, and confirming our solution from a dynamical perspective. Lower panel: Same as the upper panel, but zoomed-in on a smaller region around the $(a,e)$ of the planet, conforming its location in the quasi-stable region. 
\label{fig:megno}}
\end{figure}

We utilized the MECHANIC software\footnote{https://github.com/mslonina/Mechanic} \citep{Slonina2012a,Slonina2012b,Slonina2014} to calculate MEGNO maps for \kic, applying the latest MEGNO implementation \citep{Gozdziewski2001,Gozdziewski2003,Gozdziewski2008}. The maps have a resolution of 350 x 500 initial conditions in planetary semi-major axis $(a_p)$ and eccentricity $(e_p)$ space, each integrated for 200,000 days (corresponding to $\sim20,000$ binary periods). Quasi-periodic orbits are defined as $|\langle Y\rangle - 2.0| \simeq 0.001$; for chaotic orbits $\langle Y\rangle \rightarrow \infty$ as $t \rightarrow \infty$. The MEGNO map computed for the best-fit parameters of Table \ref{tab:pd_out} is shown in Figure \ref{fig:megno}. The cross-hair mark represents the instantaneous osculating Jacobian coordinates of \kicb. Purple (or dark) color indicates a region of quasi-periodic orbits, whereas yellow (or light) color denotes chaotic (and possibly unstable) orbits. The CBP sits comfortably in the quasi-periodic (purple) region of $(a,e)$-space between the 6:1 and 7:1 MMR \citep[not unlike Kepler-64b, see][]{kostov2013}, confirming the plausibility of our solution from a dynamical perspective. 

\section{Discussion}
\label{sec:discussion}

``{\it Why Does Nature Form Exoplanets Easily?}'', ponders \cite{heng12}. Both planetary formation scenarios of core accretion and gravitational collapse require complex processes at work and even more so for the violent environments of CBPs. Yet the plethora of discovered planets \citep{burke2013} indicates that planetary formation is ubiquitous. \cite{mart13} argue that it may be in fact easier to form planetary systems around close binary stars than around single stars, if there is a quiescent, low-turbulence layer present in the mid plane of the CB disks. Unlike disks around single stars, the surface density in a CB disk peaks in such a ``dead zone'' and, being close to the snow line, provides an ideal site for planetary formation. In addition, \cite{alex12} has shown that circumbinary disks around binary stars with $a_{bin} < 1$~AU persist longer than disks around single stars, suggesting that formation of CBPs should be commonplace. 

The $\Delta i \sim2.5\arcdeg$ misalignment of \kicb is notably larger than that of the other {\em Kepler}-discovered CBPs (with an average of $\sim0.3\arcdeg$). It is, however, comparable to the mutual inclination between Kepler-64b and its host EB, the only known quadruple stellar  system with a CBP. It is comparable to the mutual orbital inclinations of $1\arcdeg$ -- $2.3\arcdeg$ reported for the {\it Kepler} and {\it HARPS} multiplanet systems orbiting single stars, and of the Solar System value of $2.1\arcdeg$ -- $3.1\arcdeg$, including Mercury (Fabrycky et al. 2012; Fang and Margot, 2012; Figueira et al., 2012; Lissauer et al., 2011). 

\cite{quill13} argue that one plausible scenario responsible for the excitation of planetary inclinations is collisions with embryos. The authors note that measured correlations between planetary mass and inclination can provide strong clues for this scenario. While planetary masses are difficult to measure, photodynamical models of slightly misaligned CBP like \kicb can provide an important venue to test this hypothesis by providing constraints on masses and inclinations. Additionally, according to \cite{rapp13} up to 20\% of close binaries have a tertiary stellar companion, based on extrapolation from eclipse time variations (ETVs) measured for the entire \kepler~EB catalog. \cite{eggl08} find that $\sim25\%$ of all multiple systems with a solar-type star are triples and higher order. A tertiary companion on a wide orbit can be responsible for complex dynamical history of the binary system involving Kozai cycles with tidal friction \citep{kozai62,fab07b,kis98,eggl01,pej13}.

A robust correlation between occurrence rate of planets and (single) host star metallicities has been established over the past 10 years \citep{mayor11,howard13}. While it is equally likely to detect small planets around stars of wide metallicity range, giant planets \citep[$R\textgreater4R_{Earth}$, ][]{howard13} are preferentially found in orbits around metal-rich stars. Such dichotomy naturally originates from the core-accretion scenario for planet formation, with the caveat that in-situ formation may be more appropriate to describe the presence of low-mass planets close to their star \citep{howard13}. It is interesting to note that 7 of the \kepler~CB planets are gas giants, with $R\ge4.3R_{Earth}$, (the only exception being Kepler 47b) but all 7 host stellar systems are deficient in metals compared to the Sun. 

Eclipsing binary systems have long been proposed to be well-suited candidates to the discovery of transiting planets due to the favorable orbital orientation of the stellar system. However, EBs may not be as favorable as generally thought. Given the correct orientation, planets orbiting single stars will transit at every inferior conjunction. As we have shown here, and also discussed by \cite{schneider1994}, misaligned CBPs, however, may either transit or miss depending on their instantaneous orbital configuration. If the configuration is favorable, one can observe several consecutive transits. Otherwise there may be a few, widely-separated transits or even only a single transit. A trivial case is no transits at all during the course of the observations, where the planetary orbit has not yet precessed into the favorable transit geometry and the first ``good hit'' may be approaching; even a very misaligned system will occasionally transit. Thus, a non-detection of tertiary transits in the light curve of an EB does not rule out the possibility to observe a transiting CBP in the future. This statement is trivially obvious for planets with periods much longer than the duration of observations. However, as this work has illustrated, the statement also applies to short-period planetary orbits with non-zero mutual inclinations. 

Such photodynamical effects may further affect the deduced occurrence rate of CBP, even after accounting for detection efficiency, systematic effects, etc. Aligned systems have a strong selection effect, but many systems (potentially a ``silent majority'' of CBPs) could be misaligned and precessing, and \kicb will be the prototype of that class of objects. 

``{\it...The existence of planets in these systems [CBP]...}'', \cite{paar12} note, ``{\it...baffles planet formation theory...}''. The facts that the confirmed CBPs are so close to the theoretical limit for dynamical stability, and that shorter-period EBs have typically longer-period CBPs (further away from the critical limit) hint at an interesting dynamical history, and can be directly addressed by finding more CB systems. Future additions to the still small family of CBPs will add important new insight into our understanding of these remarkable objects. Or, perhaps more interestingly, the new discoveries will baffle the theoretical framework even further.

\subsection{Stellar Insolation}
\label{sec:HZ}

\begin{figure}
\centering
\plotone{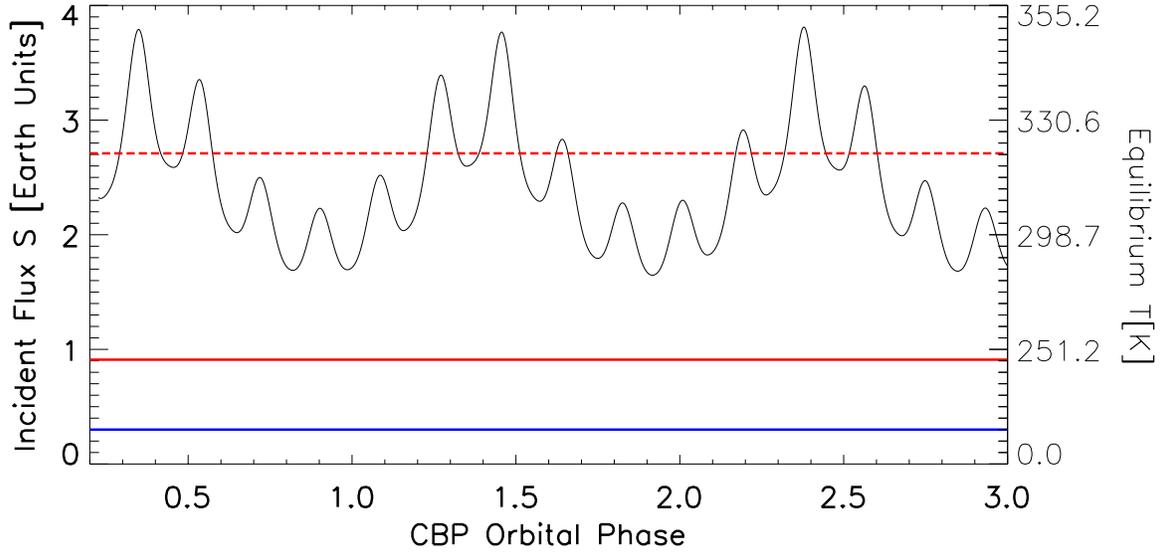}
\caption{Bolometric luminosity in units of $S_{\star} = 1368~W m^{-2}$ (the mean Solar constant), incident at the orbital location of \kicb (black line) as a function of the orbital phase of the planet, and equilibrium temperature for a Bond albedo of 0.34. The CBP orbital phase is defined as $\phi_p=t/P_p$, with $\phi_p=0$ at $t_0=2,455,014.46543$ (BJD). The planet is slightly closer to its host stars than the inner edge of the habitable zone, which is at $S_{inner}=0.91~S_{\star}$. For comparison, we show the inner (red line) and outer (blue) edges of the habitable zone of \kic. The dashed line indicates the inner edge of the dry desert habitable zone, $S_{inner, desert}=2.71~S_{\star}$; the planet is in the dry desert habitable zone for most of its orbit.
}
\label{fig:irradiance}
\end{figure}

Our best-fit photodynamical model places \kicb on a $\vaplanet$ AU-orbit around two stars with effective temperatures of $T_A = 4700$K, estimated from SOPHIE, and $T_B = 3460$K, derived from the temperature ratio $T_B/T_A$ from ELC, respectively (see Table \ref{tab_parameters}). The combined incident flux $S_{tot}=S_A+S_B$ due to the two stars A and B at the orbital location of \kicb is shown in Figure \ref{fig:irradiance}. It varies from a minimum of $\sim1.64~S_{\star}$ to a maximum of $\sim3.86~S_{\star}$ (where $S_{\star}$ is the mean Solar constant of 1368 W m$^{-2}$) on two different timescales (stellar and planetary periods), with an average of $\sim2.42~S_{\star}$. Following \cite{kane13}, we calculate the effective temperature of the EB, $T_{eff, AB}$, as that of a source with an energy flux similar to that of the two stars combined. From Wien's displacement law, and using the combined blackbody radiation of the two stars, we estimate $T_{eff, AB}\sim4500$ K. Following \cite{kopp13} cloud-free models, the inner edge of the habitable zone (``runaway greenhouse'') for the \kic system is at an incident stellar flux $S_{inner}=0.91~S_{\star}$ (red, or grey line in Figure \ref{fig:irradiance}); the outer edge (``maximum greenhouse'') is at $S_{outer}=0.28~S_{\star}$ (blue, or dark line in Figure \ref{fig:irradiance}). \kicb is slightly closer to it's host star than the inner edge of the habitable zone. We note that the inner edge distance for the habitable zone of the \kic system for dry desert planets is at $\sim0.32~AU$ (Equation 12, \cite{zsom13}), $\sim2.71~S_{\star}$, for a surface albedo of 0.2 and 1\% relative humidity. This limiting case places \kicb ($a_p=0.3553~AU$) in the dry desert habitable zone for most of its orbit.

The flux variations experienced by the CBP, coupled with the peculiar behavior of the planetary obliquity described next may result in very interesting and complex weather and climate patterns on \kicb and similar CBPs.

\subsection{Cassini States}
\label{sec:Cassini}

Next we shall discuss how the quick orbital precession, which is highly constrained by the transit fits, should affect the spin orientation of \kicb. Instantaneously, each of the stars causes a torque on the rotational bulge of the planet, but over one EB orbit, and even over one orbit of the CBP, this torque causes little reorientation of the planet. Over many orbits, however, the effect of this torque adds coherently. If we replace the stars with a point mass at their barycenter, the small-obliquity precession angular frequency of the planetary spin would be (e.g., \citealt{fab07a}):
\begin{equation}
\alpha = \frac{ k_{2,p} } { c_p } \frac{M_A+M_B}{M_p} (1-e_p^2)^{-3/2} (R_p / a_p)^3 S_p,
\end{equation}
where $k_{2,p}$ is the apsidal motion constant (half the Love number) of the CBP, $c_p$ is the normalized moment of inertia, and $S_p$ is the spin angular frequency of the planet. 

In the presence of quick orbital precession, the dynamics become much richer, as Cassini states appear \citep{ward04,ham04,fab07a,levrard07}. These states are fixed-points of the spin dynamics in which the spin and orbit precess at the same rate around the total angular momentum. Thus the effect is a 1:1 secular resonance between the orbital precession and the spin precession. The orbital precession rate, $g$, is known from the best-fitting model $g=0.57$~radians/year. Taking a 1-day rotation period (i.e. $S_p = 2\pi$~radians/day) for \kicb, $k_{2,p}=0.1$, $c_p=0.08$, and assuming $M_p=15 M_\oplus$, with the above values of the constants, we have $\alpha = 1.0$~radians/year, very close to resonant with $g$. Even precession trajectories that are not in these states are affected by them, as they must navigate the resonant island. Thus when $\alpha \approx g$, the obliquity can vary by many degrees on a timescale somewhat longer than the precession timescale. 

However, the value of $\alpha$ for the case of \kicb is very uncertain due to the poorly constrained parameters, particularly $M_p$ and $S_p$. For the best-fitting $M_p$ of $\sim60 M_\oplus$ (and other parameters as assumed above), the spin would travel around Cassini State 2 (see \cite{peale69} for the numbering), the state in which the spin responds to torques more slowly than the orbit precesses. In that case, the spin feels the precession-averaged effect of the EB orbit, and so its spin-precession pole is close to the orbit normal of the binary (the dominant angular momentum). This is the case of the Earth's moon, which has faster orbital precession than spin precession, and it is tidally damped to Cassini State 2. If \kicb has instead low mass, $M_p<10 M_\oplus$ (i.e., quite puffy), it could have a higher natural spin frequency $\alpha$. In that case, it is possible for the planet to be in a tipped-over Cassini State 2, in which a high obliquity (near $90\arcdeg$) lessens the torque from the binary star, allowing the planet precession to continue resonating with the orbital precession. However, it is more likely that it would travel around Cassini State 1, which is the normal precession around an orbit, but slightly modified due to the (relatively slow) precession of that orbit. Finally, for the Neptune-like mass of $15 M_\oplus$ assumed above, both Cassini State 1 and Cassini State 2 would be considerably displaced from the orbit normal, and either large obliquity or large obliquity fluctuations ($\sim 30\arcdeg$) would result.  

It is beyond the scope of this work to calculate the obliquity evolution of \kicb in detail. We expect, however, that it would give interesting boundary conditions for climate models \citep{lang07}. Another consideration is that the $\alpha$ value would have changed as the planet cooled, as that contraction would result in changes in $R_p$, $k_{2,p}$, $c_p$, and $S_p$; the scanning of $\alpha$ could cause trapping into a Cassini resonance \citep{winn05}. We expect that at the orbital distance of \kicb, tides would be too weak to cause spin alignment, but we note that in other systems such alignment would bring the planetary spin to a Cassini state rather than standard spin-orbit locking \citep{fab07a}.

Finally, we suggest that spin-precession of a planet may actually be observable for CBP systems. \cite{carter10} pointed out that a precessing planet will display a time-varying surface area to a transit observer, due to the oblateness of the planet changing orientations. A Saturn-like oblateness with a $30\arcdeg$ obliquity results in a few-percent change in depth over the precession cycle. The radii ratios in some CBP systems are constrained by {\em Kepler} photometry at the $\sim1\%$ level, thus variations at this level might be detectable. This is considerably more observable than the transit shape signature of oblique planets \citep{seager02,barnes03}.
\section{Conclusions}
\label{sec:conclusions}

We report the discovery of a $R_p = \vRplanet\pm\eRplanet~R_\oplus$ planet transiting the primary star of \kic. The system consists of two K+M stars that eclipse each other every 10.116146 days. Due to the small misalignment ($\Delta i\sim2.5\arcdeg$) between the binary and CBP orbital planes, the latter precesses and the planet often fails to transit the primary star. The CBP revolves around the EB every $\sim66$ days on an orbit with $a_{p} = \vaplanet$~AU and $e=\veplanet\pm\eeplanet$. The orbital configuration of the system is such that we observe a set of three transits occurring $\sim66$ days apart, followed $\sim800$ days later by five more transits also separated by $\sim66$ days from each other. We note that, among the known transiting CBPs, \kicb is the only CBP with a higher eccentricity compared to its host binary star. 

Spectroscopic measurements determined the target as a single-lined EB, and provided its mass function, eccentricity and argument of periastron. Photometric observations identified a nearby companion (``third light'') to \kic inside the central \kepler~pixel, and addressed itÕs flux contamination to the target's light curve \citep{Kostov2014b}. Based on statistical estimates, we propose that the companion star is gravitationally bound to the EB, making \kicb a CBP in a triple stellar system.

Our best-fit model places \kicb slightly closer to its host stars than the inner edge of the extended habitable zone, with the bolometric insolation at the location of the planet's orbit varying between $\sim1.75~S_{\star}$ and $\sim3.9~S_{\star}$ on multiple timescales (where $S_{\star} = 1368~W m^{-2}$, the mean Solar constant). The planet is, however in the dry desert habitable zone for most of its orbit. Also, the peculiar orbital configuration of the system indicates that \kicb may be subject to Cassini-States dynamics. Depending on the angular precession frequency of the planet, its spin and orbital precession rates could be commensurate. This suggests that \kicb may experience obliquity fluctuations of dozens of degrees on precession timescales ($\sim11$ years) and complex seasonal cycles with interesting climate patterns.  

The transits of a CBP provide precise measurements on the stellar and planetary sizes and on the masses of the host binary star. Our discovery adds to the growing knowledge about CBPs: their radii, masses, occurrence frequency about which types of stars, when they formed (first versus second generation) and even whether the concept of habitability can be extended beyond single-star planetary systems. The results reported here can be applied to studies of the formation and evolution of protoplanetary disks and planetary systems in multiple-stellar systems. 
\acknowledgments

This research was performed in partial fulfillment of the requirements of the PhD of V.B.K. at Johns Hopkins University. The authors gratefully acknowledge everyone who has contributed
to the \kepler~Mission, which is funded by NASA's Science Mission Directorate. 
We acknowledge conversations with 
Nicolas Crouzet,
Holland Ford,
K. Go{\'z}dziewski,
Nader Haghighipour,
Amy McQuillan,
Colin Norman,
Rachel Osten,
Neill Reid,
Jean Schneider,
M. S{\l{}}onina,
and
Martin Still. The authors thank the referee for the helpful comments and suggestions.

This research used observations made with the SOPHIE instrument on the 1.93-m telescope at Observatoire de Haute-Provence (CNRS), France, as part of programs 12B.PNP.MOUT and 13A.PNP.MOUT. 
This research made use of the the SIMBAD database, operated at CDS, Strasbourg, France;
data products from the Two Micron All Sky Survey (2MASS),
the Digitized Sky Survey (DSS),
the NASA exoplanet archive NexSci\footnote{http://exoplanetarchive.ipac.caltech.edu};
source code for transit light curves (Mandel and Agol 2002);
SFI/HEA Irish Centre for High-End Computing (ICHEC);
Numerical computations presented in this work were partly  carried out using the SFI/HEA Irish Centre for High-End Computing (ICHEC, STOKES) and the PLUTO computing cluster at KASI;
Astronomical research at Armagh Observatory is funded by the Department of Culture,  Arts and Leisure (DCAL).
V.B.K. and P.R.M. received funding from NASA Origins of Solar Systems grant NNX10AG30G, and NESSF grant NNX13AM33H.
W.F.W. and J.A.O. gratefully acknowledge support from the National Science Foundation via grant AST-1109928, and from NASA's \kepler~Participating Scientist Program (NNX12AD23G) and Origins of Solar Systems Program
(NNX13AI76G).
T.C.H acknowledges support by the Korea Research Council for Science and Technology (KRCF) through the Young Scientist Research Fellowship Program grant number 2013-9-400-00. T.M. acknowledges support from the European Research Council under the EU's Seventh Framework Programme (FP7/(2007-2013)/ ERC Grant Agreement No.~291352) and from the Israel Science Foundation (grant No.~1423/11).
\bibliographystyle{apj}

\newpage

\clearpage
\begin{table}[ht]
\begin{center}
\footnotesize
\caption{Parameters of the Binary Star Systems.
\label{tab_parameters}}
\begin{tabular}{llllll}
\hline
\hline
{\bf \kic} & & & & &  \\
\hline
Parameter & Symbol & Value & Uncertainty (1$\sigma$) & Unit & Note \\
\hline
Orbital Period & $P_{bin}$ & 10.116146 & 0.000001 & d & {\it Kepler} photometry \\
Epoch of primary eclipse & $T_{prim}$ & 2454972.981520 & - & BJD & Pr{\v s}a et al. (2011) \\
Epoch of secondary eclipse & $T_{sec}$ & 2454977.999 & 0.001 & BJD &  Pr{\v s}a et al. (2011) \\
Epoch of Periastron passage & $T_{0}$ & 2454973.230 & 0.023 & BJD &  SOPHIE \\
Velocity semi-amplitude & $K_1$ & 43.489 & 0.085 & km s$^{-1}$ & SOPHIE \\
Velocity offset & $\gamma$ & -27.78& 0.05 & km s$^{-1}$ & SOPHIE \\
Argument of Periapsis & $\omega_{bin}$ & 279.54$^\dagger$$^\dagger$ & 0.86 & \arcdeg & SOPHIE \\
Eccentricity   & $e_{bin}$ & 0.0372 & 0.0017 & & SOPHIE \\
Orbital Inclination & $i_{bin}$ & 87.3258 & 0.0987 & \arcdeg & {\it Kepler} photometry$^\dagger$ \\
Normalized Semimajor Axis & $a_{bin}/R_A$ & 27.5438 & 0.0003 & & {\it Kepler} photometry$^\dagger$ \\
Fractional Radius & $R_B/R_A$ & 0.5832 & 0.0695 & & {\it Kepler} photometry$^\dagger$ \\
Temperature of Star A & $T_A$ & 4700 & - & K & Spectroscopic \\
Temperature ratio & $T_B/T_A$ & 0.7369 & 0.0153 & K & {\it Kepler} photometry$^\dagger$ \\
Limb-Darkening Coeff. of Star A & $x_A$ & 0.3567 & 0.0615 & K & {\it Kepler} photometry$^\dagger$ \\
V sin i of Star A & $V sin i$ & 5 & 2 & km s$^{-1}$ & SOPHIE \\
Fe/H of Star A & $[Fe/H]$ & -0.2 & - & & NexSci$^\dagger$$^\dagger$$^\dagger$ \\
Gravity of Star A & $\Logg_A$ & 4.67 & - & & NexSci \\
\hline
\hline
\multicolumn{4}{l}{$\dagger$: ELC fit \citep{Orosz2012}} \\
\multicolumn{4}{l}{$\dagger$$\dagger$: Observer at -z} \\
\multicolumn{4}{l}{$\dagger$$\dagger$$\dagger$: http://exoplanetarchive.ipac.caltech.edu} \\
\end{tabular}
\end{center}
\end{table}

\clearpage
\begin{table}[ht]
\begin{center}
\caption{Measured radial velocities.
\label{tab_RV}}
\begin{tabular}{lrr}
\hline
\hline
BJD$_{\rm UTC}$ & RV & $\pm$$1\,\sigma$ \\
$-2\,400\,000$ & (km\,s$^{-1}$) & (km\,s$^{-1}$)\\
\hline
56\,180.42595 & $-59.89$ & 0.20  \\
56\,184.39404 & 15.59 & 0.19  \\
56\,186.44561 & $-14.52$ & 0.09  \\
56\,187.47375 & $-43.73$ & 0.10  \\
56\,192.42495 & $-16.21$ & 0.16  \\
56\,195.40345$\dagger$ & 9.99 & 0.25  \\
56\,213.36121 & $-0.37$ & 0.17  \\
56\,362.67907 & $-56.59$ & 0.16  \\
56\,401.55880 & $-71.26$ & 0.14  \\
56\,403.56461 & $-47.72$ & 0.27  \\
56\,404.62680 & $-21.87$ & 0.19  \\
\hline
\end{tabular}
\\
$\dagger$: measurement corrected for sky background pollution.
\end{center}
\end{table}

\clearpage
\begin{table}[ht]
\begin{center}
\label{pd_in}
\footnotesize
\caption{Model parameters for the photometric-dynamical model. We adopt the ``best-fit'' values as the system's parameters. The reference epoch is $t_0 =2,455,014.46543$ (BJD).
\label{tab:pd_in}}
\begin{tabular}{|ll|llll|}
\hline
Index & Parameter Name & Best-fit & 50\% & 15.8\% & 84.2\% \\ \hline
&~{\it Mass parameters} & & & & \\
0&~~RV Semi-Amplitude Star A, $K_A$ (km s$^{-1}$) & $                                   43.42$ & $                                   43.49 $ & $-                                    0.16$ & $+                                    0.19$\\
1&~~Mass ratio, Star B, $M_B/M_A$ & $                                  0.6611$ & $                                  0.6592 $ & $-                                  0.0035$ & $+                                  0.0034$\\
2&~~Planetary mass ratio, $M_p/M_A$ ($\times 1000$) & $                                   0.245$ & $                                   0.186 $ & $-                                   0.078$ & $+                                   0.078$\\
&~{\it Stellar Orbit} & & & &\\
3&~~Orbital Period, $P_{bin}$ (day) & $                              10.1161114$ & $                              10.1161185 $ & $-                               0.0000101$ & $+                               0.0000099$\\
4&~~Time of Barycentric Passage, $t_{bin}-2455000$ (BJD) & $                                 8.34898$ & $                                 8.34902 $ & $-                                 0.00024$ & $+                                 0.00024$\\
5&~~Eccentricity Parameter, ${e_{bin}} \sin(\omega_{bin})$ & $                                 -0.0359$ & $                                 -0.0360 $ & $-                                  0.0023$ & $+                                  0.0022$\\
6&~~Eccentricity Parameter, ${e_{bin}} \cos(\omega_{bin})$ & $                                0.006169$ & $                                0.006166 $ & $-                                0.000037$ & $+                                0.000038$\\
7&~~Orbital Inclination, $i_{bin}$ (deg) & $                                  87.332$ & $                                  87.301 $ & $-                                   0.060$ & $+                                   0.050$\\
&~{\it Planetary Orbit} & & & &\\
8&~~Orbital Period, $P_p$ (day) & $                                  66.262$ & $                                  66.269 $ & $-                                   0.021$ & $+                                   0.024$\\
9&~~Time of Barycenteric Passage, $t_p-2455000$ (BJD) & $                                   96.64$ & $                                   96.57 $ & $-                                    0.17$ & $+                                    0.16$\\
10&~~Eccentricity Parameter, $\sqrt{e_p} \sin(\omega_p)$ & $                                  0.3426$ & $                                  0.3435 $ & $-                                  0.0033$ & $+                                  0.0031$\\
11&~~Eccentricity Parameter, $\sqrt{e_p} \cos(\omega_p)$ & $                                  -0.027$ & $                                  -0.022 $ & $-                                   0.013$ & $+                                   0.014$\\
12&~~Orbital Inclination, $i_p$ (deg) & $                                  89.929$ & $                                  89.942 $ & $-                                   0.016$ & $+                                   0.024$\\
13&~~Relative Nodal Longitude, $\Delta \Omega_p$ (deg) & $                                   3.139$ & $                                   3.169 $ & $-                                   0.064$ & $+                                   0.080$\\
&~{\it Radius/Light Parameters} & & & &\\
14&~~Linear Limb Darkening Parameter, $u_A$ & $                                   0.599$ & $                                   0.643 $ & $-                                   0.036$ & $+                                   0.036$\\
15&~~Density of Star A, $\rho_A$ (g cm$^{-3}$) & $                                   1.755$ & $                                   1.799 $ & $-                                   0.049$ & $+                                   0.066$\\
16&~~Radius Ratio, Star B, $R_B/R_A$ & $                                   0.624$ & $                                   0.650 $ & $-                                   0.032$ & $+                                   0.043$\\
17&~~Planetary Radius Ratio, $R_p/R_A$ & $                                  0.0514$ & $                                  0.0517 $ & $-                                  0.0013$ & $+                                  0.0013$\\
18&~~Stellar Flux Ratio, $F_B/F_A$ ($\times 100$) & $                                    5.90$ & $                                    6.40 $ & $-                                    0.76$ & $+                                    1.05$\\
&~{\it Relative Contamination, $F_{\rm cont}/F_A$} ($\times 100$) & & & &\\
19&~~All Seasons & $                                     7.6$ & $                                     8.0 $ & $-                                     1.0$ & $+                                     1.0$\\
&~{\it Noise Parameter} & & & & \\
20&~~Long Cadence Relative Width, $\sigma_{\rm LC}$ ($\times 10^5$) & $                                   67.78$ & $                                   67.76 $ & $-                                    0.53$ & $+                                    0.54$\\
&~{\it Radial Velocity Parameters} & & & & \\
21&~~RV Offset, $\gamma$ (km s$^{-1}$) & $                                 -27.784$ & $                                 -27.810 $ & $-                                   0.113$ & $+                                   0.098$\\
22&~~RV Jitter, $\sigma_{\rm RV}$ (km s$^{-1}$) & $                                    0.01$ & $                                    0.17 $ & $-                                    0.11$ & $+                                    0.20$\\
\hline
\end{tabular}
\end{center}
\end{table}

\clearpage
\begin{table}[ht]
\begin{center}
\caption{Derived parameters from the photometric-dynamic model. We adopt the ``best-fit'' values as the system's parameters. The reference epoch is $t_0 =2,455,014.46543$ (BJD).
\label{tab:pd_out}}
\begin{tabular}{|l|llll|}
\hline
~~Parameter & Best-fit & 50\% & 15.8\% & 84.2\% \\ \hline
~{\it Bulk Properties} & & & & \\
~~Mass of Star A, $M_A$ ($M_\odot$) & $                                   0.820$ & $                                   0.830 $ & $-                                   0.014$ & $+                                   0.015$\\
~~Mass of Star B, $M_B$ ($M_\odot$) & $                                  0.5423$ & $                                  0.5472 $ & $-                                  0.0073$ & $+                                  0.0081$\\
~~Mass of Planet b, $M_p$ ($M_\oplus$) & $                                     67.$ & $                                     51. $ & $-                                     21.$ & $+                                     22.$\\
~~Radius of Star A, $R_A$ ($R_\odot$) & $                                  0.7761$ & $                                  0.7725 $ & $-                                  0.0096$ & $+                                  0.0088$\\
~~Radius of Star B, $R_B$ ($R_\odot$) & $                                   0.484$ & $                                   0.502 $ & $-                                   0.021$ & $+                                   0.027$\\
~~Radius of Planet p, $R_p$ ($R_\oplus$) & $                                   4.347$ & $                                   4.352 $ & $-                                   0.099$ & $+                                   0.099$\\
~~Density of Star A, $\rho_A$ (g cm$^{-3}$) & $                                   1.755$ & $                                   1.799 $ & $-                                   0.049$ & $+                                   0.066$\\
~~Density of Star B, $\rho_B$ (g cm$^{-3}$) & $                                    4.77$ & $                                    4.32 $ & $-                                    0.63$ & $+                                    0.58$\\
~~Density of Planet, $\rho_p$ (g cm$^{-3}$) & $                                     3.2$ & $                                     2.4 $ & $-                                     1.0$ & $+                                     1.0$\\
~~Gravity of Star A, $\log g_A$ (cgs) & $                                  4.5721$ & $                                  4.5811 $ & $-                                  0.0086$ & $+                                  0.0108$\\
~~Gravity of Star B, $\log g_B$ (cgs) & $                                   4.802$ & $                                   4.774 $ & $-                                   0.046$ & $+                                   0.036$\\
~{\it Orbital Properties} & & & & \\
~~Semimajor Axis of Stellar Orbit, $a_{bin}$ (AU) & $                                 0.10148$ & $                                 0.10185 $ & $-                                 0.00052$ & $+                                 0.00057$\\
~~Semimajor Axis of Planet, $a_p$ (AU) & $                                  0.3553$ & $                                  0.3566 $ & $-                                  0.0018$ & $+                                  0.0020$\\
~~Eccentricity of Stellar Orbit, $e_{bin}$ & $                                  0.0365$ & $                                  0.0366 $ & $-                                  0.0021$ & $+                                  0.0023$\\
~~Argument of Periapse Stellar Orbit, $\omega_{bin}$ (Degrees) & $                                  279.74$ & $                                  279.71 $ & $-                                    0.58$ & $+                                    0.62$\\
~~Eccentricity of Planetary Orbit , $e_p$ & $                                  0.1181$ & $                                  0.1185 $ & $-                                  0.0017$ & $+                                  0.0018$\\
~~Argument of Periapse Planet Orbit, $\omega_p$ (Degrees) & $                                    94.6$ & $                                    93.6 $ & $-                                     2.3$ & $+                                     2.2$\\
~~Mutual Orbital Inclination, $\Delta i$ (deg)$^\dagger$ & $                                   4.073$ & $                                   4.121 $ & $-                                   0.083$ & $+                                   0.113$\\
\hline
\multicolumn{4}{l}{$\dagger$: $\cos(\Delta i) = \sin(i_{bin})\sin(i_p)\cos(\Delta\Omega) + \cos(i_{bin})\cos(i_p)$} \\
\end{tabular}
\end{center}
\end{table}

\begin{table}[ht]
\begin{center}
\scriptsize
\caption{Mid-transit times, depths and durations of the planetary transits.
\label{tab:future}}
\begin{tabular}{lllllll||ll}
\hline
\hline
 Event \# & Center & $\sigma$ & Depth$^\dagger$ & $\sigma$ & Duration & $\sigma$ & Center & Duration \\
 & (Time-2455000 [BJD]) & (Center) & [ppm] & (Depth) & [days] & (Duration) & (Time-2455000 [BJD]) & [days] \\
\hline
\hline
{\bf Observed} & & & & & & & {\bf Predicted} & \\
\hline
1 & -4.3799 & 0.0019 & 1557 & 668 & 0.1517 & 0.0113 & -4.38 & 0.14 \\
2 & 62.3363 & 0.0018 & 2134 & 537 & 0.18 & 0.0138 & 62.34 & 0.18 \\
3 & 125.0938 & 0.0033 & 2958 & 678 & 0.1549 & 0.0145 & 125.1 & 0.16 \\
-- & -- & -- & -- & -- & -- & -- & 188.34$^\dagger$$^\dagger$ & 0.1 \\
4 & 963.1529 & 0.0045 & 2662 & 406 & 0.1551 & 0.0209 & 963.16 & 0.16 \\
5 & 1026.1544 & 0.0037 & 2376 & 381 & 0.1083 & 0.0062 & 1026.16 & 0.12 \\
6 & 1092.3978 & 0.0075 & 2759 & 322 & 0.3587 & 0.0199 & 1092.40 & 0.36 \\
7 & 1156.2889 & 0.0057 & 1892 & 394 & 0.0921 & 0.0144 & 1156.29 & 0.1 \\
8 & 1219.5674 & 0.0084 & 3282 & 432 & 0.2149 & 0.0236 & 1219.56 & 0.22 \\
\hline
{\bf Future} & & & & & & \\
\hline
9 & -- & -- & -- & -- & -- & -- & 3999.47 & 0.12 \\
\hline
\multicolumn{4}{l}{$\dagger$: in terms of $(\frac{r_p}{r_A})^2$} \\
\multicolumn{4}{l}{$\dagger$$\dagger$: Predicted transit, difficult to be detected in the data} \\
\end{tabular}
%}
\end{center}
\end{table}
\end{document}